\documentclass[
  journal=pasa,
  manuscript=research-paper, 
  year=2022,
  volume=??,
]{cup-journal}

\usepackage{microtype,siunitx,booktabs}
\usepackage{natbib}
\setcitestyle{aysep={}}

\usepackage{CJKutf8}    % Chinese characters
\usepackage{tikz,xcolor,hyperref} % For ORCID
\usepackage{amsmath}    % For multi-line equations
\usepackage{amssymb}    % For lesssim
\usepackage{longtable}
\usepackage{multicol}
\usepackage{float}

\definecolor{lime}{HTML}{A6CE39}
\DeclareRobustCommand{\orcidicon}{%
    \begin{tikzpicture}
    \draw[lime, fill=lime] (0,0) 
    circle [radius=0.16] 
    node[white] {{\fontfamily{qag}\selectfont \tiny ID}};
    \draw[white, fill=white] (-0.0625,0.095) 
    circle [radius=0.007];
    \end{tikzpicture}
    \hspace{-1mm}
}

\newcommand{\orcidChrisO}{\href{https://orcid.org/0000-0003-0017-349X}{\orcidicon}}
\newcommand{\orcidChrisW}{\href{https://orcid.org/0000-0002-4569-016X}{\orcidicon}}
\newcommand{\orcidSamuel}{\href{https://orcid.org/0000-0001-9372-4611}{\orcidicon}}
\newcommand{\orcidPatrick}{\href{https://orcid.org/0000-0003-4237-0520}{\orcidicon}}
\newcommand{\orcidJack}{\href{https://orcid.org/0000-0001-8359-2328}{\orcidicon}}
\newcommand{\orcidRachel}{\href{https://orcid.org/0000-0002-5325-2709}{\orcidicon}}

\newcommand\ion[2]{\text{#1\,\textsc{\lowercase{#2}}}}	% ionization states
\newcommand\arcsec{$^{\prime\prime}$}
\newcommand\arcmin{$^{\prime}$}

\title{AllBRICQS: the All-sky BRIght, Complete Quasar Survey}

\author{{Christopher A.\ Onken}~\orcidChrisO}
\affiliation{
Research School of Astronomy and Astrophysics, Australian National University, Canberra ACT 2611, Australia}
\email[Christopher A.\ Onken]{christopher.onken@anu.edu.au}

\author{~Christian Wolf~\orcidChrisW}
\affiliation{
Research School of Astronomy and Astrophysics, Australian National University, Canberra ACT 2611, Australia}
\alsoaffiliation{
Centre for Gravitational Astrophysics, Australian National University, Canberra ACT 2600, Australia}

\author{~Wei Jeat Hon~\orcidJack}
\affiliation{ 
School of Physics, University of Melbourne, Parkville, VIC 3010, Australia}

\author{~Samuel Lai~\begin{CJK}{UTF8}{gbsn}(赖民希)\end{CJK}~\orcidSamuel}
\affiliation{ 
Research School of Astronomy and Astrophysics, Australian National University, Canberra ACT 2611, Australia}

\author{~Patrick Tisserand~\orcidPatrick}
\affiliation{ 
Sorbonne Universit\'{e}s, UPMC Univ Paris 6 et CNRS, Institut d'Astrophysique de Paris, 98 bis bd Arago, F-75014 Paris, France}

\author{~Rachel Webster~\orcidRachel}
\affiliation{
School of Physics, University of Melbourne, Parkville, VIC 3010, Australia}

\doi{}

\received {18 Sep 2022}
\revised  {dd Mmm YYYY}
\accepted {dd Mmm YYYY}
\published{dd Mmm YYYY}

 %% First letter not capped

\begin{document}

\begin{abstract}
We describe the first results from the All-sky BRIght, Complete Quasar Survey (AllBRICQS), which aims to discover the last remaining optically bright quasars.
We present 156 spectroscopically confirmed quasars (140 newly identified) having $|b|>10^{\circ}$. 152 of the quasars have {\it Gaia} DR3 magnitudes brighter than $B_{P}=16.5$ or $R_{P}=16$~mag, while four are slightly fainter.
The quasars span a redshift range of $z=0.07-3.93$.
In particular, we highlight the properties of J0529-4351 at $z=3.93$, which, if unlensed, is one of the most intrinsically luminous quasars in the Universe.
The AllBRICQS sources have been selected by combining data from the {\it Gaia} and {\it WISE} all-sky satellite missions, and we successfully identify quasars not flagged as candidates by {\it Gaia} Data Release 3. We expect the completeness to be $\approx96$\% within our magnitude and latitude limits, while the preliminary results indicate a selection purity of $\approx96$\%. The optical spectroscopy used for source classification will also enable detailed quasar characterisation, including black hole mass measurements and identification of foreground absorption systems. The AllBRICQS sources will greatly enhance the number of quasars available for high-signal-to-noise follow-up with present and future facilities.

\end{abstract}

\section{Introduction}

Optically bright quasars represent an admixture of sources drawn from a continuum of properties --- at one extreme: the rare, distant, rapidly growing black holes (BHs); and at the other: the more common, modestly growing BHs located closer to the Milky Way. Previous work has found the cumulative surface density of optically bright quasars to be well described by a power-law as a function of magnitude, such that the integrated number of quasars brighter than $B_{J}$=15.5~mag is 1 per 100~deg$^{2}$, and increases by 1~dex for every 1.3~magnitudes of decreased optical flux \citep{2000A&A...358...77W}. 

Compared to the bulk of quasars at fainter optical magnitudes, the bright end of the distribution offers obvious advantages in terms of signal-to-noise for detailed follow-up. Such quasars are particularly useful as background sources for studies of intervening absorption from circumgalactic \citep{2017ARA&A..55..389T} and intergalactic gas \citep{2016ARA&A..54..313M}. 

Moreover, because the physical size of the quasar broad-line region (BLR) scales with luminosity as $\sim \sqrt{L}$ \citep{2013ApJ...767..149B}, and, to first order, the distance to the quasar scales as $\sqrt{L/F}$ for flux $F$, the {\it angular} size of the BLR scales as $\sqrt{F}$. This means that the optically brightest quasars have the largest apparent BLRs, making them especially interesting targets for high spatial resolution investigation with instruments such as GRAVITY on the Very Large Telescope Interferometer \cite[VLTI; e.g.,][]{2018Natur.563..657G, 2020A&A...643A.154G, 2021A&A...648A.117G, 2021A&A...654A..85G}.

Approximately 1\% of spectroscopically confirmed active galactic nuclei (AGNs)\footnote{These statistics are based on the Million Quasars Catalogue \citep{2021arXiv210512985F}, version 7.6; hereafter, Milliquas.} have {\it Gaia} Data Release 3 photometry \citep{2016A&A...595A...1G, 2021A&A...649A...1G} brighter than $B_{P}=16.5$ or $R_{P}=16$~mag, and 89\% of those bright objects have redshifts less than $z=0.1$. The recent discovery of a remarkably bright quasar at $z=0.83$ --- SMSS~J114447.77-430859.3 (hereafter, J1144-4308), with $B_{P}=14.64$~mag \citep{2022arXiv220604204O} --- motivates a renewed search for other bright quasars that have gone undetected.

Here we describe the search criteria and first results from the All-sky BRIght, Complete Quasar Survey (AllBRICQS). In Section~\ref{sec:selection}, we explain the AllBRICQS selection criteria. In Section~\ref{sec:obs}, we describe the optical spectroscopy obtained for the first set of candidates. We present the spectroscopically confirmed AllBRICQS sample in Section~\ref{sec:sample} and report on our preliminary success rate. In Section~\ref{sec:J0529}, we examine the highest-redshift AllBRICQS discovery, J0529-4351, in greater detail. Section~\ref{sec:Gaia_comparison} compares our confirmed quasars with the quasar catalogues derived from {\it Gaia} DR3. In Section~\ref{sec:discussion}, we discuss the present results and the outlook for the scientific exploitation of the completed survey.

We adopt a flat $\Lambda$CDM cosmology, having a matter density of $\Omega_{\rm m}=0.3$ and a Hubble-Lema\^itre constant of $H_0$ = 70~km~s$^{-1}$~Mpc$^{-1}$. Throughout the paper, we use Vega magnitudes unless otherwise specified.

\section{Candidate Selection}\label{sec:selection}

\begin{figure*}[t]
\centering
\vspace{0pt}\includegraphics[width=0.351\textwidth]{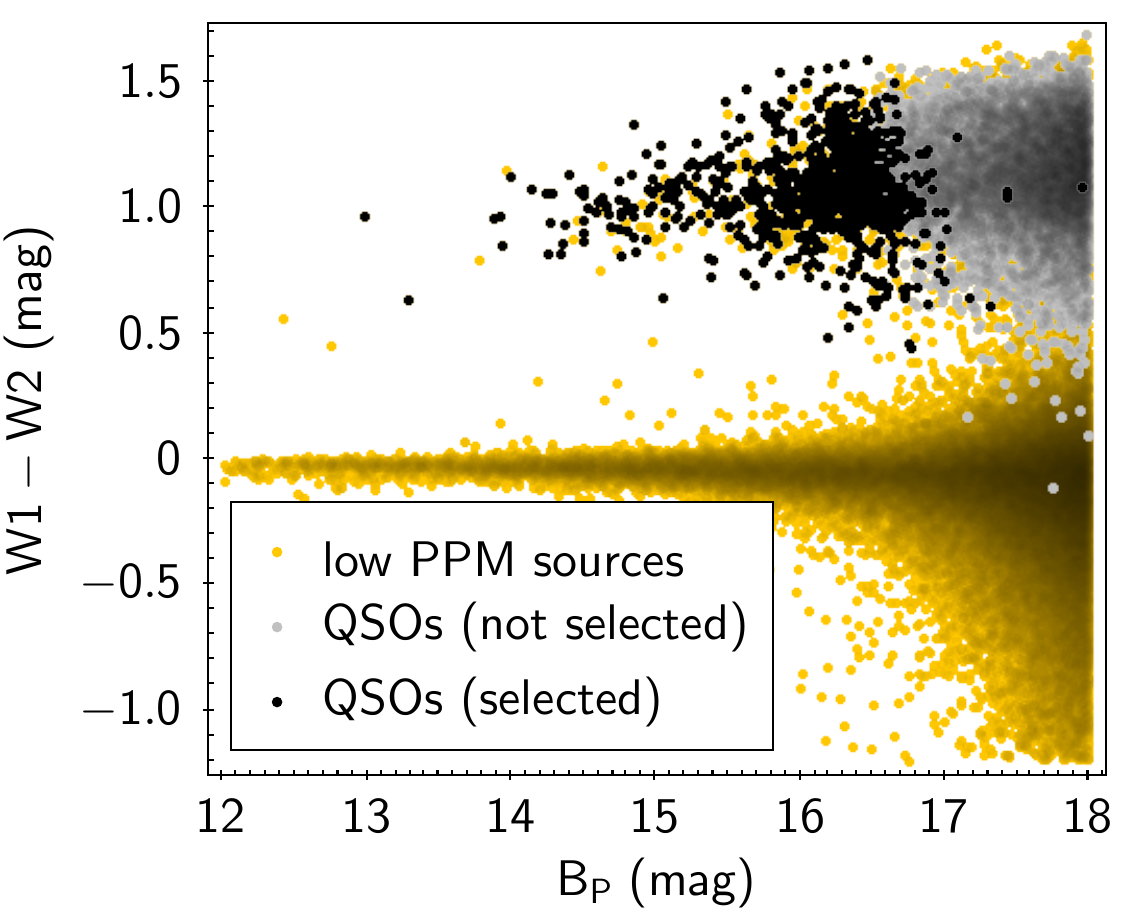}
\vspace{0pt}\includegraphics[width=0.324\textwidth]{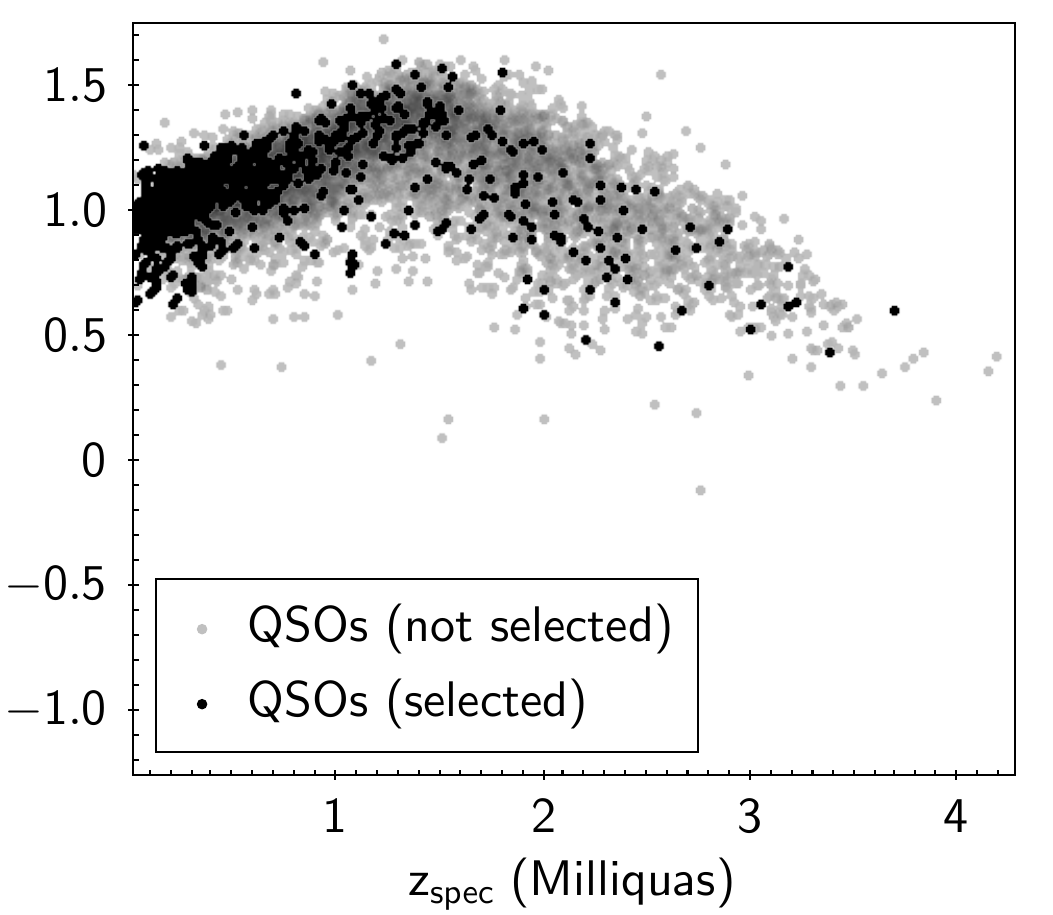}
\vspace{0pt}\includegraphics[width=0.318\textwidth]{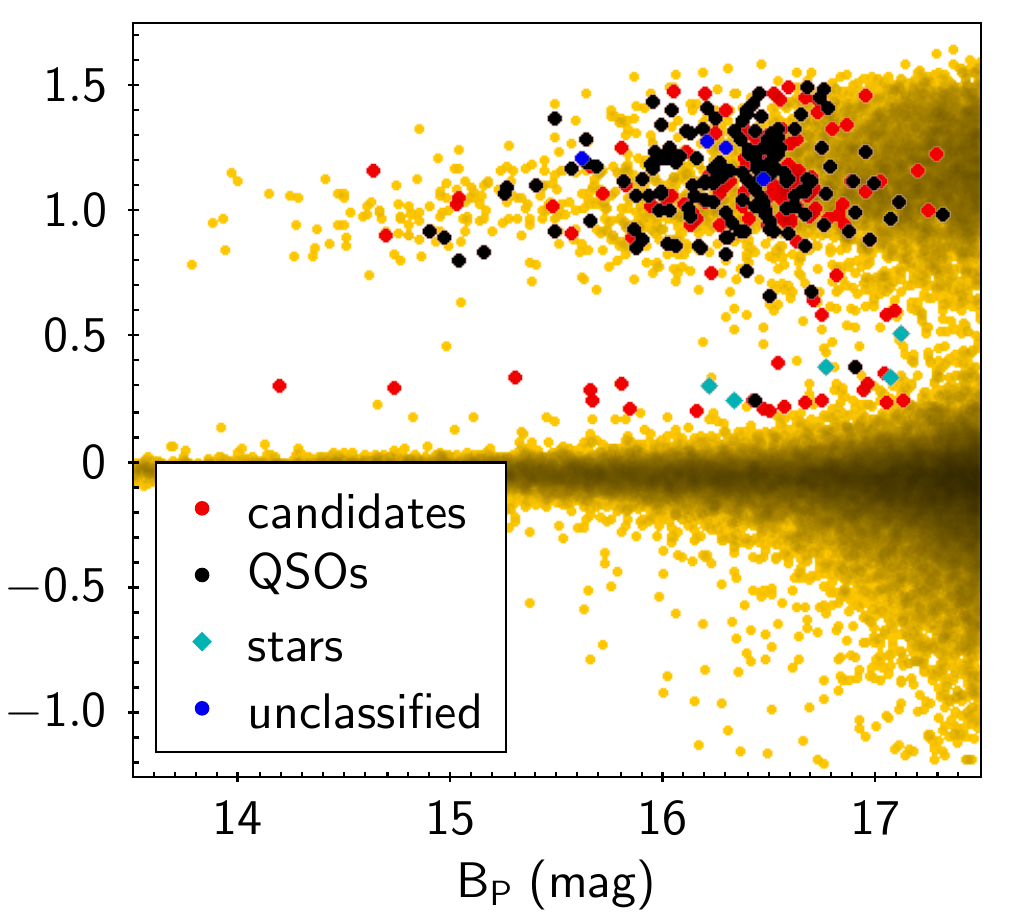}
\caption{
Left panel: {\it WISE} $W1-W2$ colour versus {\it Gaia} $B_{P}$ magnitude of objects with low parallax and proper motion (yellow); known quasars are shown in black or, if they violate our selection criterion, in grey. Centre panel: $W1-W2$ colour versus redshift for known quasars satisfying (black) or violating (grey) our selection criteria. Right panel: $W1-W2$ colour versus $B_{P}$ magnitude for objects with low PPM in yellow, the new AllBRICQS quasars in black, contaminating stars in cyan, unclassified spectra in blue, and unobserved candidates in red.
}
\label{fig:w1m2_qsos}
\end{figure*}

Finding bright quasars has been greatly simplified in modern times, first with the all-sky survey of the {\it Wide-field Infrared Survey Explorer} \cite[{\it WISE;}][]{2010AJ....140.1868W,2014ApJ...792...30M}, and more recently  with the accurate parallax and proper motion information from the all-sky survey of {\it Gaia}. Several authors have proposed very effective methods for selecting quasar candidates exploiting these data sets using hard selection cuts, Bayesian probabilities, or machine-learning approaches \citep[e.g.][]{2019ApJ...887..268C, 2019MNRAS.489.4741S, 2020MNRAS.491.1970W}. Here, we choose a very simple approach, given that we focus our interest at bright magnitudes, where the stars, galaxies, and quasars can be easily separated, as demonstrated below. 

We avoid the vicinity of the Galactic Plane, where quasars are dimmed by dust extinction and are harder to identify reliably due to source crowding biasing the photometry, especially in the {\it WISE} catalogue, where the large point spread function in the infrared passbands leads to blending. Effective searches in the Galactic Plane include the work of \citet{2022ApJS..261...32F}, who have recently discovered 191 new quasars, five of which are brighter than $i=16$~ABmag. In contrast, we restrict our search to a Galactic latitude of $|b|>10^\circ$ and avoid the most difficult parts of the plane.

Previous work has shown that known quasars have {\it Gaia} parallaxes and proper motions (PPM) that are statistically consistent with zero within their errors \citep[e.g.,][but for interesting counter-examples, see \citealp{2022FrASS...922768W}]{2021PASP..133i4501L,2022MNRAS.511..572O,2022A&A...660A..16S} and hence we restrict the search to a target list with parallaxes and proper motions of low ($<4\sigma$) significance. We also restrict the search to objects with a {\it Gaia} $B_{P}/R_{P}$ excess factor \cite[phot\_bp\_rp\_excess\_factor; see][]{2021A&A...649A...3R} $<1.5$, which rejects clearly extended or blended sources. Among the known broad-line quasars in the Milliquas catalogue having {\it Gaia} $B_{P}<17$~mag, $z>0.3$, and $|b|>10^{\circ}$, the selection rules above reject less than 4\% of the quasars (43 out of 1,227); some of the rejected objects have no PPM measurements at all, however, the vast majority of the objects are simply consistent with being drawn from the tails of a zero-PPM population. 

We investigated several colour properties, and found that the mid-infrared $W1-W2$ colour provides a highly effective distinction between the known quasars and the bulk of the candidate sample, which are primarily distant (hence, small PPM) stars. In the left panel of Figure~\ref{fig:w1m2_qsos}, we show the $W1-W2$ colour vs. {\it Gaia} $B_{P}$ magnitude for the known quasars from Milliquas and the sample passing the {\it Gaia} criteria above. At $B_{P}<17$~mag, the quasar and non-quasar population appear clearly separated, while they start to blend into each other at fainter magnitudes. As the known quasars have been compiled from a variety of selection methods (X-ray, UV, optical, radio), the mid-infrared quasar colours should not be subject to any particular biases beyond the scatter arising from the photometric errors, and just reflect the typical spectral energy distribution (SED) associated with accretion onto a BH. Thus, we select quasar candidates using the simple rule of $W1-W2>0.2$~mag. While there is likely to be increasing contamination towards that $W1-W2$ boundary, the moderate number of candidates to follow up allows us to pursue a strategy that emphasises completeness (and the threshold may be relaxed further in future expansions of the observing campaign). Based on the PPM statistics above, expect the completeness to be $\approx96$\% for $|b|>10^{\circ}$.

As shown in the middle panel of Figure~\ref{fig:w1m2_qsos}, the quasars that approach our $W1-W2$ threshold at faint $B_{P}$ magnitudes are strongly biased towards high redshift, but some sources at $z\sim 4$ can still be found within our selected region.

We limit the present sample to have either $B_{P}<16.5$ or $R_{P}<16$~mag and find 290 candidates\footnote{An early version of the sample definition extended the depth a further 0.5~magnitudes, and we present four quasars discovered from that sample in Tables~\ref{tab:sample} and \ref{tab:lum}, as well as the spectrum gallery in \ref{app:gallery}.} (see right panel of Figure~\ref{fig:w1m2_qsos}). Approximately 500 broad-line quasars are known at $z>0.3$ within these magnitude limits. If $\sim 85$\% of our candidates were similar objects, then one third of bright quasars would be missing from current catalogues and be revealed by AllBRICQS. 

\begin{figure*}
\centering
\includegraphics[width=0.33\textwidth]{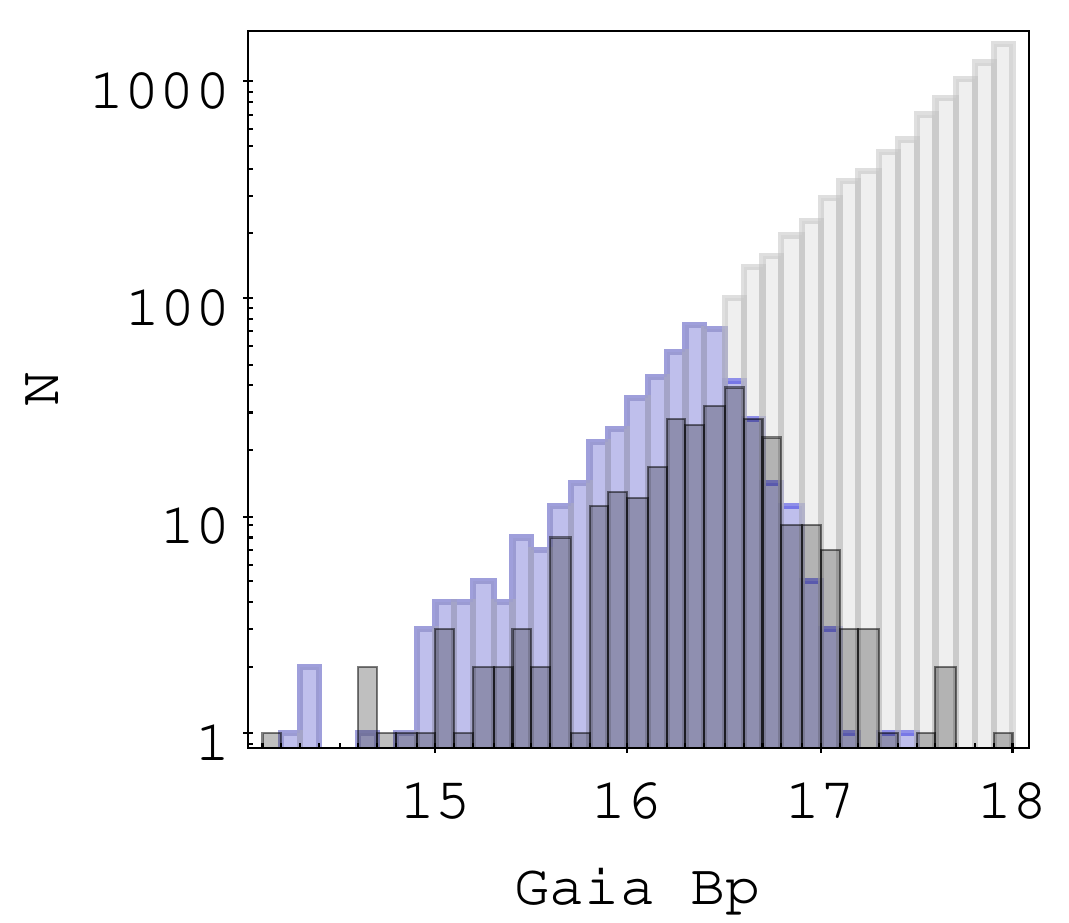}
\includegraphics[width=0.33\textwidth]{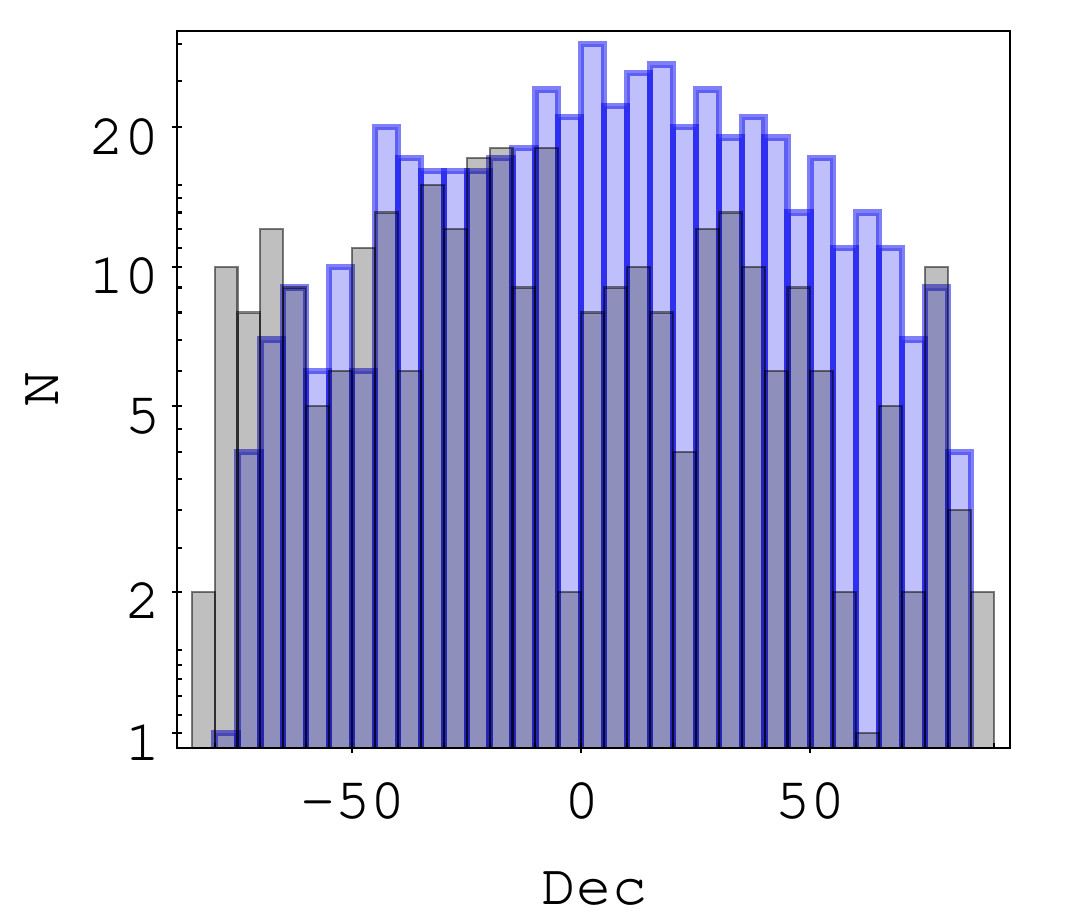}
\includegraphics[width=0.33\textwidth]{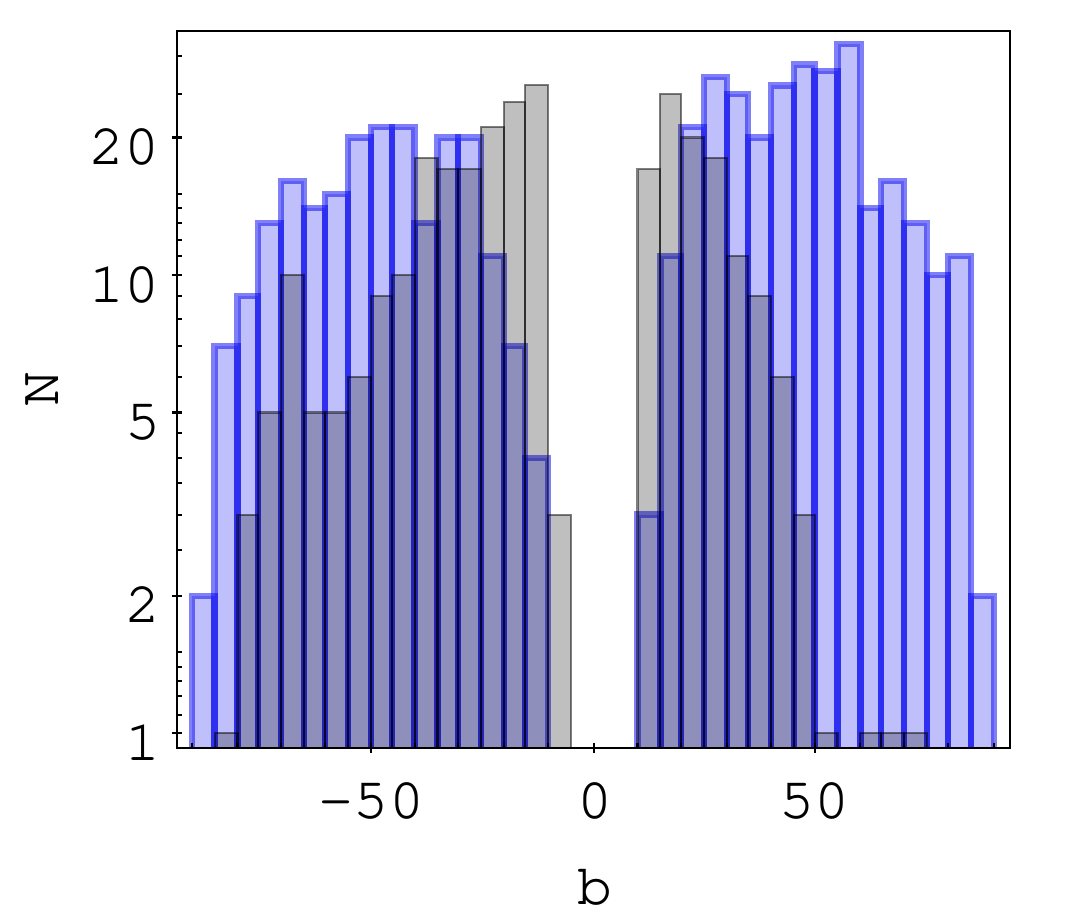}
\caption{Number counts of AllBRICQS candidates (dark grey) as a function of {\it Gaia} $B_{P}$ magnitude (left), declination (centre), and Galactic latitude (right). The blue histograms are known quasars for the same magnitude range as the candidates: {\it Gaia} $B_{P}<16.5$ or $R_{P}<16$~mag. Light grey are all known quasars.}
\label{fig:hist_known_cands}
\end{figure*}

In Fig.~\ref{fig:hist_known_cands} we characterise our candidate list further in relation to known quasars of similar brightness. We can see that our new candidates are more numerous at Southern declinations and especially at Southern Galactic latitudes. The reason is, of course, that the Sloan Digital Sky Survey \citep[SDSS,][]{2000AJ....120.1579Y,2002AJ....123.2945R,2022ApJS..259...35A}, which has revealed a huge sample of quasars, was focused on the Northern Galactic Cap.

\section{Observations and Data Processing}\label{sec:obs}

Optical spectroscopy was obtained with the Australian National University (ANU) 2.3-meter telescope at Siding Spring Observatory (SSO) using the Wide Field Spectrograph \cite[WiFeS;][]{2007Ap&SS.310..255D,2010Ap&SS.327..245D}. WiFeS is an integral field spectrograph and, with a spatial binning of 2 pixels, provides $1x1$\arcsec sampling over a $25x38$\arcsec field-of-view. 

WiFeS spectra were obtained between UT~2022-03-11 and 2022-11-24 for the AllBRICQS sample presented here. We note that observations on one of those nights (UT~2022-06-14) were conducted as part of the Science Verification activities for the fully robotic operations of the ANU 2.3m telescope. The targets were selected from the candidate list based purely on considerations of observability (Right Ascension and airmass), aiming for complete coverage of the list (between RA=10h and 5h, only five candidates still needed observing).

With the resolving power $R\sim3000$ gratings and the RT560 beamsplitter, we used exposure times between 180 and 1500~s, covering the wavelength range $3250-9550$~\AA\ across the two cameras of the spectrograph. Various spectrophotometric standard stars were observed alongside the AllBRICQS candidates and were used for calibrating the overall spectral shape. However, because the standards were not necessarily observed under the same observing conditions, we do not utilise them for absolute flux calibration.

The raw data were processed with the {\sc Python}-based pipeline, PyWiFeS \citep{2014Ap&SS.349..617C}. We then extracted the spectra from the 3D data cubes (separate cubes for each arm of the spectrograph) using {\sc QFitsView}\footnote{Available from \url{https://www.mpe.mpg.de/~ott/QFitsView/}.}, utilising nearby regions for sky subtraction. The signal-to-noise values in the final spectra were typically between 10 and 50 per pixel.

The spectra were classified and redshifted by eye, aided by the {\sc Marz} online application\footnote{See \url{http://samreay.github.io/Marz}.} \citep{2016A&C....15...61H,2016ascl.soft05001H}. The typical redshift uncertainty is 0.01, but can be either better or worse depending on the details of the available emission lines and their shapes.

\section{AllBRICQS Sample}\label{sec:sample}

The observations reported here encompass 166 of the 290 AllBRICQS candidates. We present a list of the 156 confirmed AllBRICQS quasars in \ref{app:sample}, arranged in order of ascending RA. A gallery of the quasar spectra, arranged in order of ascending redshift, is provided in \ref{app:gallery}. Four unclassified sources are presented in \ref{app:unknown}. Finally, the six stars we observed are presented in \ref{app:nonqso}. 

As noted in \ref{app:sample}, 140 of our 156 quasars represent new discoveries, while the remaining 16 are composed of quasars identified from the Edinburgh-Cape Blue Object Survey \cite[13 quasars without published redshifts, and so intentionally retained in the candidate list;][]{1997MNRAS.287..848S,2016MNRAS.459.4343K}; two quasars of known redshift, but lacking published optical spectroscopy; and one quasar announced since the beginning of the AllBRICQS observing campaign. In contrast, just six candidates have proven to be stars. Excluding the four quasars beyond our {\it Gaia} magnitude limits and the four unclassified sources, this gives a preliminary estimate of the AllBRICQS selection purity of 152/158 = 96\%. 

The right panel of Figure~\ref{fig:w1m2_qsos} suggests that the purity is declining at the blue end of the $W1-W2$ range, but we do find quasars near the boundary and will wait for the completed sample before making a more comprehensive statistical analysis.

\subsection{Luminosities}

We flux-calibrate the WiFeS spectra for the AllBRICQS quasars with the {\it Gaia} DR3 $B_{P}$ and $R_{P}$ photometry and then correct for Galactic extinction using the \citet{2019ApJ...886..108F} extinction curve implemented in the \texttt{dust\_extinction} {\sc Python} package \citep{2021zndo...4658887G}. We assume $R_V=3.1$, and apply a $\times0.86$ correction factor to the values drawn from the \citet{1998ApJ...500..525S} $E(B-V)$ map \citep{2011ApJ...737..103S}. We derive continuum luminosities from the observed redshifts and the median extinction-corrected emission in 20~\AA\ windows centred around the rest-frame wavelengths of 1450, 3000, and 5100~\AA\ (as the redshifts permit, and excluding the 1450~\AA\ continuum for broad absorption line quasars). We also estimate the bolometric luminosity for each quasar by applying the bolometric corrections (BCs) from \citet{2012MNRAS.422..478R, 2012MNRAS.427.1800R} and adopting the mean when multiple values are available. The continuum and bolometric luminosities are provided in \ref{app:lum}. Based on the typical RMS fluxes within the spectroscopic continuum windows, we expect the luminosity estimates to be reliable at the level of 0.1~dex, but plan to make more precise measurements when performing detailed spectral fitting in future work.

\begin{figure*}[!ht]
\centering
\includegraphics[width=\textwidth]{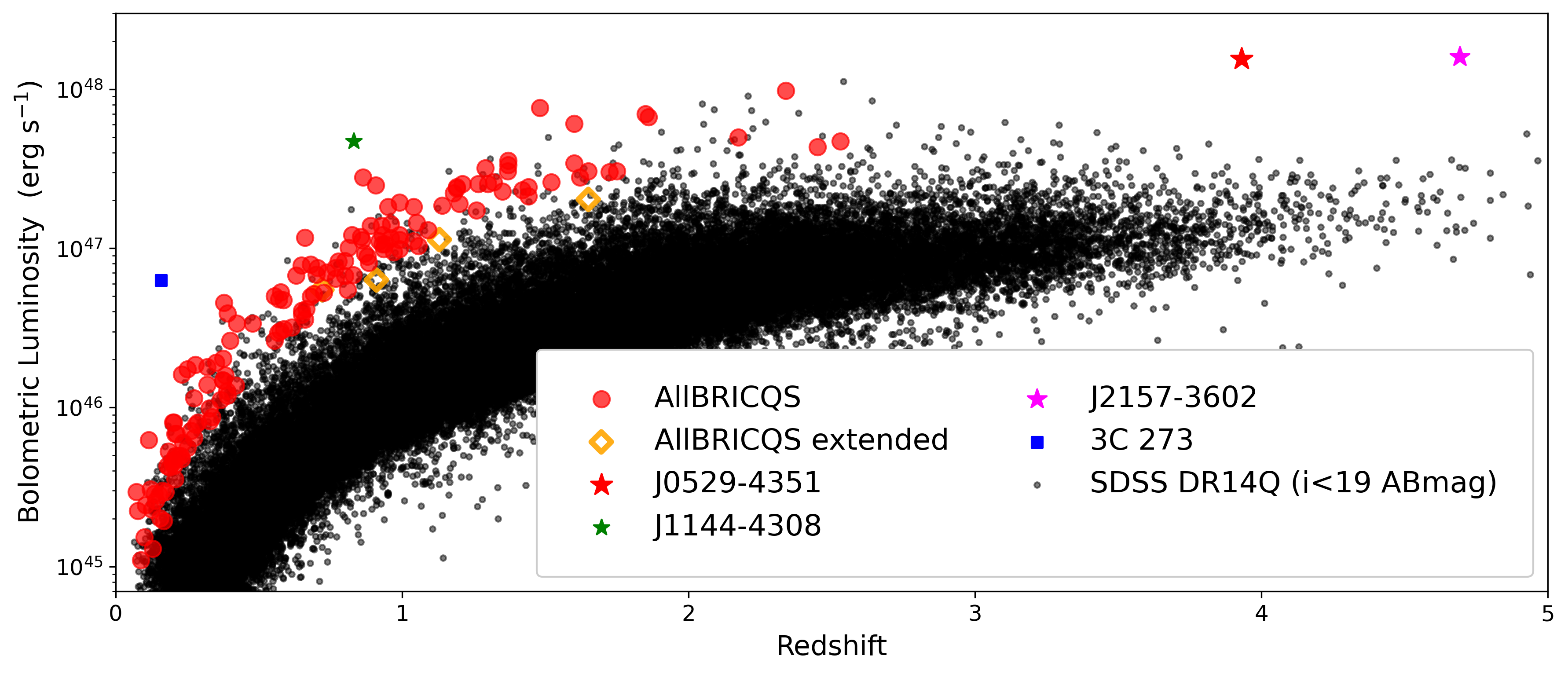}
\caption{Bolometric luminosity versus redshift for the first AllBRICQS quasars (red circles), with the highest-redshift source (J0529-4351; see Sect.~\ref{sec:J0529}) indicated as a red star. The AllBRICQS discoveries that are fainter than $B_{P}=16.5$ and $R_{P}=16$~mag are indicated with open orange diamonds. For comparison, we also show 3C~273 (blue square); the recently discovered J1144-4308 \cite[green star;][]{2022arXiv220604204O}; the most luminous known quasar, J2157-3602 \cite[magenta star;][]{2018PASA...35...24W,2020MNRAS.496.2309O}; and the numerous quasars observed by SDSS \cite[grey points;][]{2020ApJS..249...17R}. The AllBRICQS quasars improve our knowledge of the luminous end of the quasar distribution across a wide redshift range.}
\label{fig:zl}
\end{figure*}

The bolometric luminosities across redshift are shown in comparison to the quasars from the 14th data release of SDSS \cite[DR14Q;][]{2020ApJS..249...17R} in Figure~\ref{fig:zl}. To adopt a consistent bolometric correction as above, we apply the \citet{2012MNRAS.422..478R, 2012MNRAS.427.1800R} BCs to the tabulated DR14Q continuum luminosities and again take the mean value when multiple estimates are available. We also show in Figure~\ref{fig:zl} a number of bright quasars for comparison with the AllBRICQS sample: the long-studied 3C~273; the most luminous known quasar in the Universe, SMSS~J215728.21-360215.1 \cite[hereafter, J2157-3602;][]{2018PASA...35...24W,2020MNRAS.496.2309O}; and the bright quasar which inspired the AllBRICQS project, J1144-4308 \citep{2022arXiv220604204O}. The AllBRICQS quasars fill a niche not well populated by the DR14Q sample, despite substantial overlap between the SDSS bright-end limit of $i_{\rm SDSS}=15$~ABmag and the AllBRICQS faint-end limit of $R_{P}=16$~mag ($i_{\rm SDSS}\approx 16.35$~ABmag).

For the extended AllBRICQS sample (having {\it Gaia} magnitudes just fainter than our main sample's criteria of $B_{P}<16.5$~mag or $R_{P}<16$~mag), we find one quasar, J1501-1053 at $z=0.72$, with a luminosity that places it amongst the bulk of the AllBRICQS population ($5.5\times10^{46}$~erg~s$^{-1}$). This quasar is relatively blue ($B_{P}-R_{P}$=0.59~mag), yet is in the 74th percentile of reddening amongst our confirmed quasars, with an $E(B-V)_{\rm SFD}$=0.12~mag. The extinction-corrected continuum luminosities at both 3000~\AA\ and 5100~\AA\ provide a consistent estimate of the bolometric luminosity, indicating that it was the foreground dust that caused the source to be marginally outside our selection criteria.

\subsection{Gravitational Lensing}\label{sec:lensing}

One mechanism for producing high apparent luminosities is gravitational lensing, but we do not find evidence that the AllBRICQS sample is significantly affected by such magnifications. Here, we describe the general findings for the AllBRICQS sample, with particular discussion of the $z=3.93$ quasar, J0529-4351, deferred to Section~\ref{sec:J0529}.

First, we have visually inspected the optical images available from the SkyMapper Southern Survey\footnote{See \url{https://skymapper.anu.edu.au}.} \citep{2018PASA...35...10W,2019PASA...36...33O}, the Pan-STARRS1 Survey\footnote{See \url{https://ps1images.stsci.edu/cgi-bin/ps1cutouts}.} \citep{2016arXiv161205560C,2020ApJS..251....4W}, and the Dark Energy Camera (DECam) archive\footnote{See \url{https://datalab.noirlab.edu/sia.php}.} \citep{2015AJ....150..150F,2014ASPC..485..379V}, and find no indications of small-separation point sources that would be consistent with being multiple, lensed images of the quasar (although the low-redshift quasars often have visible host galaxies, including three --- J0400-2257, J1619-7832, and J2126-4529
--- which appear to be hosted by one component of a major galaxy merger). 

Furthermore, we have examined the properties of the AllBRICQS quasars in the catalogues of the DECam Local Volume Exploration Survey (DELVE) DR2 \citep{2022ApJS..261...38D} and the NOIRLab Source Catalog (NSC) DR2 \citep{2021AJ....161..192N}. In neither catalogue do we find any AllBRICQS sources at $z>0.5$ with significant evidence of being spatially extended. 

Taking advantage of the spatial resolution of {\it Gaia}, we also computed the $C^{*}$ statistic of \citet{2021A&A...649A...3R}, which corrects the phot\_bp\_rp\_excess\_factor for colour-based trends. The $C^{*}$ values compare the fluxes in the wide-aperture spectroscopic windows ($B_{P}$ and $R_{P}$) with the narrow-aperture astrometric window ($G$-band), and are constructed to be 0 for point sources. Comparing the $C^{*}$ values with their expected ($G$-magnitude-based) uncertainties, we find the vast majority of high $C^{*}/\sigma$ quasars are the low redshift quasars in which the host galaxy is expected to contribute significant flux. At $z>0.5$, only 9\% (9 of 96 quasars) have $C^{*}/\sigma>2$, and 2\% (2 quasars: J0117-1712 at $z=0.68$, and J0530+0042 at $z=0.7$) have $C^{*}/\sigma>3$. For the $>3\sigma$ quasars, the former has deep imaging by the Dark Energy Survey \citep{2021ApJS..255...20A} that shows a source 1.5\arcsec\ away. However, the {\it Gaia} DR3 astrometry for this $G=19.71$~mag neighbour reveals it to be a foreground star. Meanwhile, J0530+0042 has DECam imaging (NOIRLab Prop. ID 2014B-0193; PI: F.~Walter), which shows a faint source ($\sim20.5$~ABmag in $i-$ and $z$-band), 3.5\arcsec\ to the Northeast, which is not catalogued by {\it Gaia} DR3, DELVE DR2, or NSC DR2. As this neighbour is beyond the 3.5-arcsec-wide $B_{P}/R_{P}$ aperture, the relation to the high $C^{*}$ value is unclear, but we conclude that there is no convincing evidence for a foreground lens.

The lack of multiply imaged quasars is likely a consequence of our selection criteria. Requiring phot\_bp\_rp\_excess\_factor $<1.5$ would eliminate $\sim65$\% of the known gravitational lenses from either the Gravitationally Lensed Quasar Database\footnote{See \url{https://research.ast.cam.ac.uk/lensedquasars/}.} \cite[see][]{2019MNRAS.483.4242L} or from a recent compilation of lensed, binary, and projected quasar pairs \citep{2022arXiv220607714L}. Additional lensing systems are likely to be lost from the AllBRICQS candidate list by violation of our $W1-W2$ criterion through the mid-IR contributions from the lensing galaxy.

Thus, we find that the AllBRICQS quasars presented here are intrinsically luminous sources, drawn from the bright end of the quasar luminosity function.

\section{The Case of J0529-4351}\label{sec:J0529}

In this section, we focus on the properties of the $z=3.93$ quasar, J0529-4351. As indicated in Figure~\ref{fig:zl}, the continuum flux at rest-frame 1450~\AA\ suggests an intrinsic bolometric luminosity of $1.5\times10^{48}$~erg~s$^{-1}$, on par with the most luminous known quasar, J2157-3602 \citep{2018PASA...35...24W, 2020MNRAS.496.2309O}. In the observed frame, aside from two gravitationally lensed systems (APM~08279+5255 and B~1422+231), J0529-4351 has the brightest {\it Gaia} $R_{P}$ magnitude of any quasar more distant than HS~1946+7658 at $z=3.051$. Hence, we examine in detail the available imaging data to constrain the possibility of gravitational lensing boosting the observed flux.

As discussed in Section~\ref{sec:lensing}, the relative flux measured by {\it Gaia} through its narrow ($G$-band) and wide ($B_{P}$ and $R_{P}$) apertures can be used as a measure of extended emission on sub-arcsecond spatial scales. For J0529-4351, the colour-corrected phot\_bp\_rp\_excess\_factor, $C^{*}$, is found to be -0.037. With the 1$\sigma$ scatter expected to be 0.02 at $G=16.34$~mag \citep{2021A&A...649A...3R}, the {\it Gaia} data do not indicate any additional flux at radii larger than 0.175\arcsec from the quasar (half the 0.35\arcsec aperture of the $G$-band). 
 
In addition, the Dark Energy Survey (DES) DR2 \citep{2021ApJS..255...20A} photometry for J0529-4351 is found to be consistent with a point source in all five optical/infrared bands ($grizY$): the \texttt{SPREAD\_MODEL} and \texttt{WAVG\_SPREAD\_MODEL} are below 0.001 \cite[where values below 0.003 have been found to robustly isolate stellar sources in $griz$ imaging of similar depth;][]{2012ApJ...757...83D}, and the \texttt{CLASS\_STAR} estimates are $>0.98$.

To assess possible microlensing of J0529-4351 from stars in a putative foreground galaxy, we examined the photometry from several surveys of the past decade. SMSS DR3 optical photometry\footnote{SMSS DR3 is currently available to Australian astronomers, but the upcoming SMSS DR4 will be rapidly made available to astronomers worldwide.} indicates a dimming of $\sim0.2$~mag between November 2014 and October 2019 in $griz$ (the source is undetected in the $u$ and $v$ filters because of the Ly$\alpha$ absorption). 
Similarly, the photometry from the NEOWISE 2022 Data Release\footnote{See \url{https://wise2.ipac.caltech.edu/docs/release/neowise/}.} \citep{2014ApJ...792...30M} in $W1$ and $W2$, obtained between March 2014 and September 2021, indicates no more than 0.05~mag of dimming over that time span.

In terms of large-separation lensed images, the only other {\it Gaia} DR3 source in a 30\arcmin\ search radius having similar colours, low PPM, and an $R_{P}$ within 2~magnitudes of J0529-4351 is shown by DES imaging to be the centre of a foreground spiral galaxy.

We find that the existing data for J0529-4351 provides no evidence that the quasar is gravitationally lensed, and thus, we conclude that its extraordinary luminosity is likely to be an indication of an extremely high accretion rate onto its BH. 

In terms of its multi-wavelength properties, we find no radio detections within 2\arcmin\ in DR1 of the Rapid ASKAP Continuum Survey \cite[RACS;][]{2020PASA...37...48M,2021PASA...38...58H}, and no X-ray counterparts catalogued within 25\arcmin\ in the Second {\it ROSAT} all-sky X-ray Survey \cite[2RXS;][]{2016A&A...588A.103B}.

Given the unusual nature of J0529-4351, we have already begun to collect additional data to further investigate this remarkable quasar.

\section{Comparison with {\it Gaia}-selected quasars}\label{sec:Gaia_comparison}

We compare our confirmed quasars with the quasar catalogues derived from the {\it Gaia} variability analysis \citep{2022arXiv220706849C}, or from other {\it Gaia} source classification methods \citep{2022arXiv220605681G,2022arXiv220606710D}, including those utilising the $B_{P}/R_{P}$ low-resolution spectra \citep{2022arXiv220606143D,2022arXiv220606205M}. 

First, we consider the {\it Gaia} DR3 variabLE AgN sample \cite[GLEAN;][]{2022arXiv220706849C}. The GLEAN coverage excludes the Galactic Plane and regions around the Magellanic Clouds because of its constraint on the number density of sources within 100~arcsec being less than 0.004~arcsec$^{-2}$. This irregular exclusion zone covers $|b|\lesssim25^{\circ}$ at the Galactic Centre, but allows sparse sampling of the Plane in the direction of the Anti-centre. The GLEAN sample\footnote{See \url{https://gea.esac.esa.int/archive/documentation/GDR3/Data_analysis/chap_cu7var/sec_cu7var_agn/}.} totals over 872,000 sources, including more than 21,000 potentially new quasars. The latter are predominantly at $G_{P}>20$~mag, significantly fainter than the AllBRICQS regime ($G_{P}\lesssim17$~mag).

We indicate in Table~\ref{tab:sample} whether each AllBRICQS quasar was included in the GLEAN sample, or in another {\it Gaia} DR3 variability table\footnote{See \url{https://gea.esac.esa.int/archive/documentation/GDR3/Data_analysis/chap_cu7var/sec_cu7var_intro/ssec_cu7var_dataproducts.html}.}, or neither. The completeness of the GLEAN sample is estimated \cite[cf. Fig.~24 of][]{2022arXiv220706849C} to be $>80$\% for $G_{P}<17$~mag, with a purity of $>95$\%. Having 123 of our 156 quasars selected by GLEAN, and one GLEAN-selected star compared to the 123 quasars, the first AllBRICQS sample is consistent with the \citet{2022arXiv220706849C} estimates.

We further compare the AllBRICQS quasars to the classifications produced by {\it Gaia}'s Discrete Source Classifier \cite[DSC;][]{2022arXiv220605681G,2022arXiv220606710D}, a probabilistic classification algorithm that was trained on 300,000 SDSS quasars. The DSC assigned a quasar label (via the Combmod classifier, recorded in the classlabel\_dsc parameter) to 116 of our 156 quasars, or $74$\% completeness within the present sample. This appears to be significantly lower than the expected\footnote{See Fig.~11.13 from \url{https://gea.esac.esa.int/archive/documentation/GDR3/Data_analysis/chap_cu8par/sec_cu8par_apsis/ssec_cu8par_apsis_dsc.html}.} quasar completeness for $G_{P}<17$~mag of nearly 100\%. We find that 37 of the AllBRICQS quasars were labelled as stars (including one as a "physicalbinary"), as was just 1 of the 6 stars we observed. Of particular note, J0529-4351, the luminous quasar at $z=3.93$, was labelled as a star by the DSC with 99.98\% probability.

In addition to the 116 quasars identified by the DSC, another 10 were flagged as quasars by the Outlier Analysis (OA) module, which uses self-organising maps \citep{kohonen-self-organized-formation-1982} to label sources with low DSC classification probability.

\begin{figure*}
\centering
\includegraphics[width=0.45\textwidth]{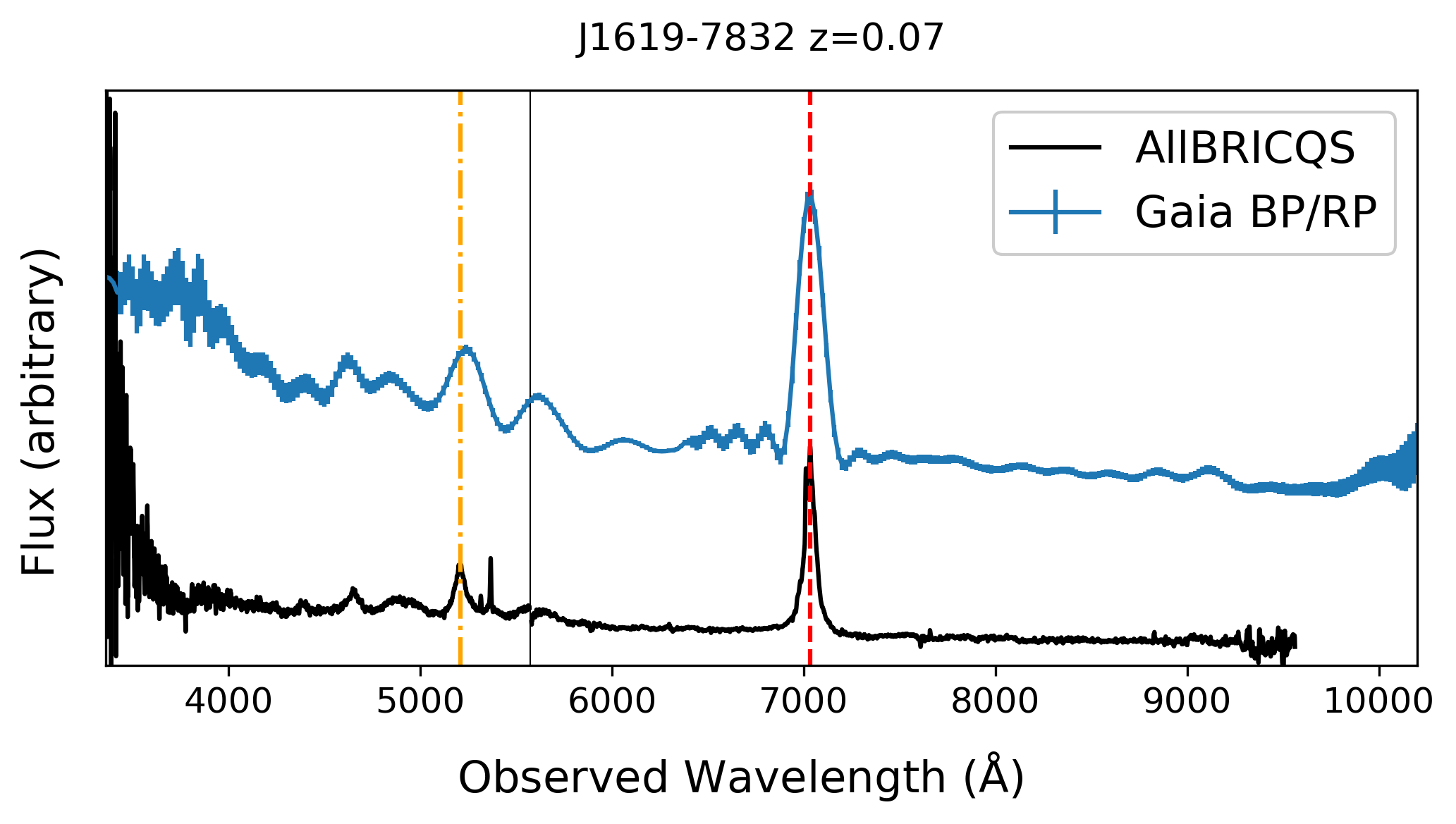}
\includegraphics[width=0.45\textwidth]{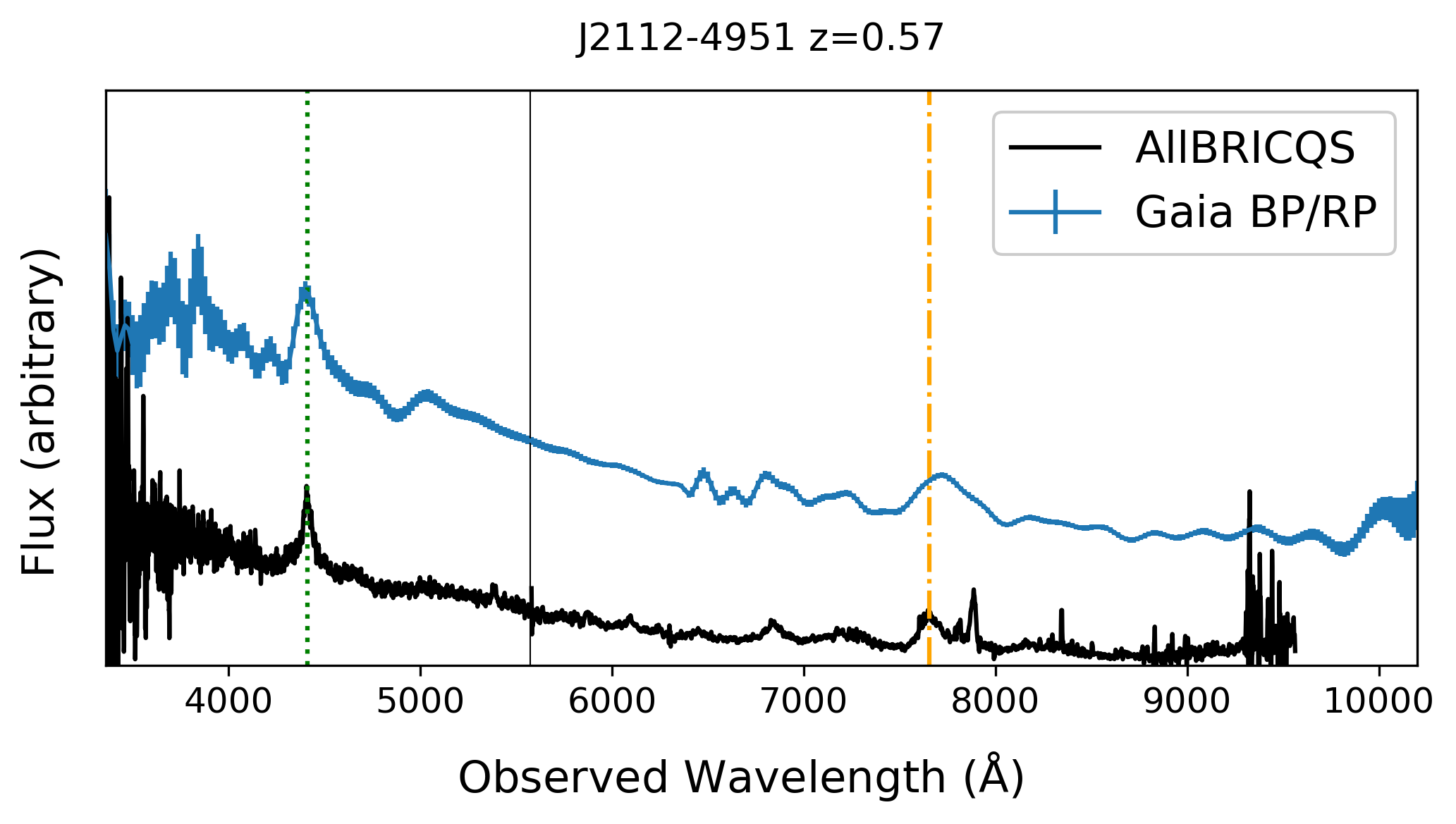}\\
\includegraphics[width=0.45\textwidth]{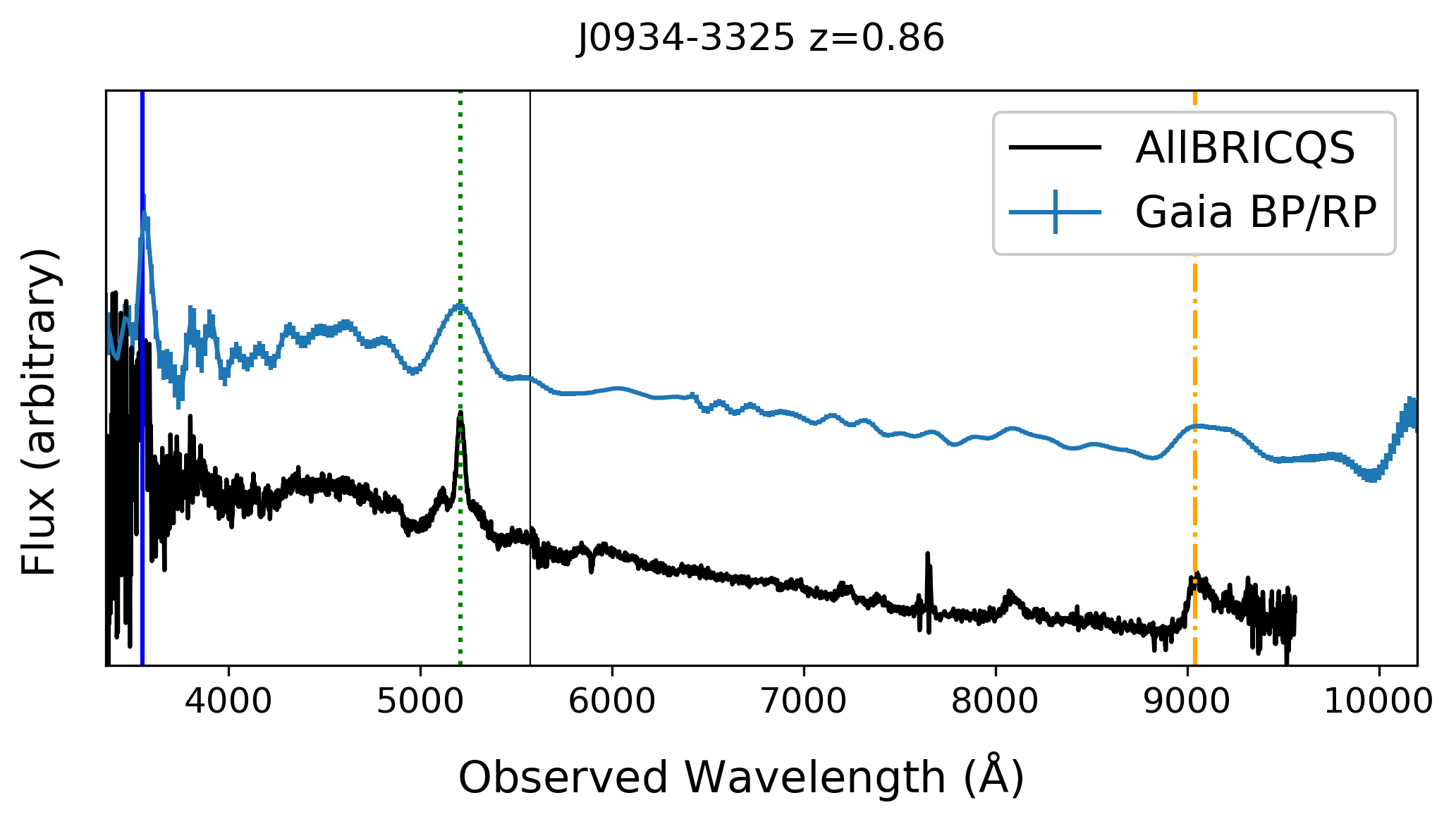}
\includegraphics[width=0.45\textwidth]{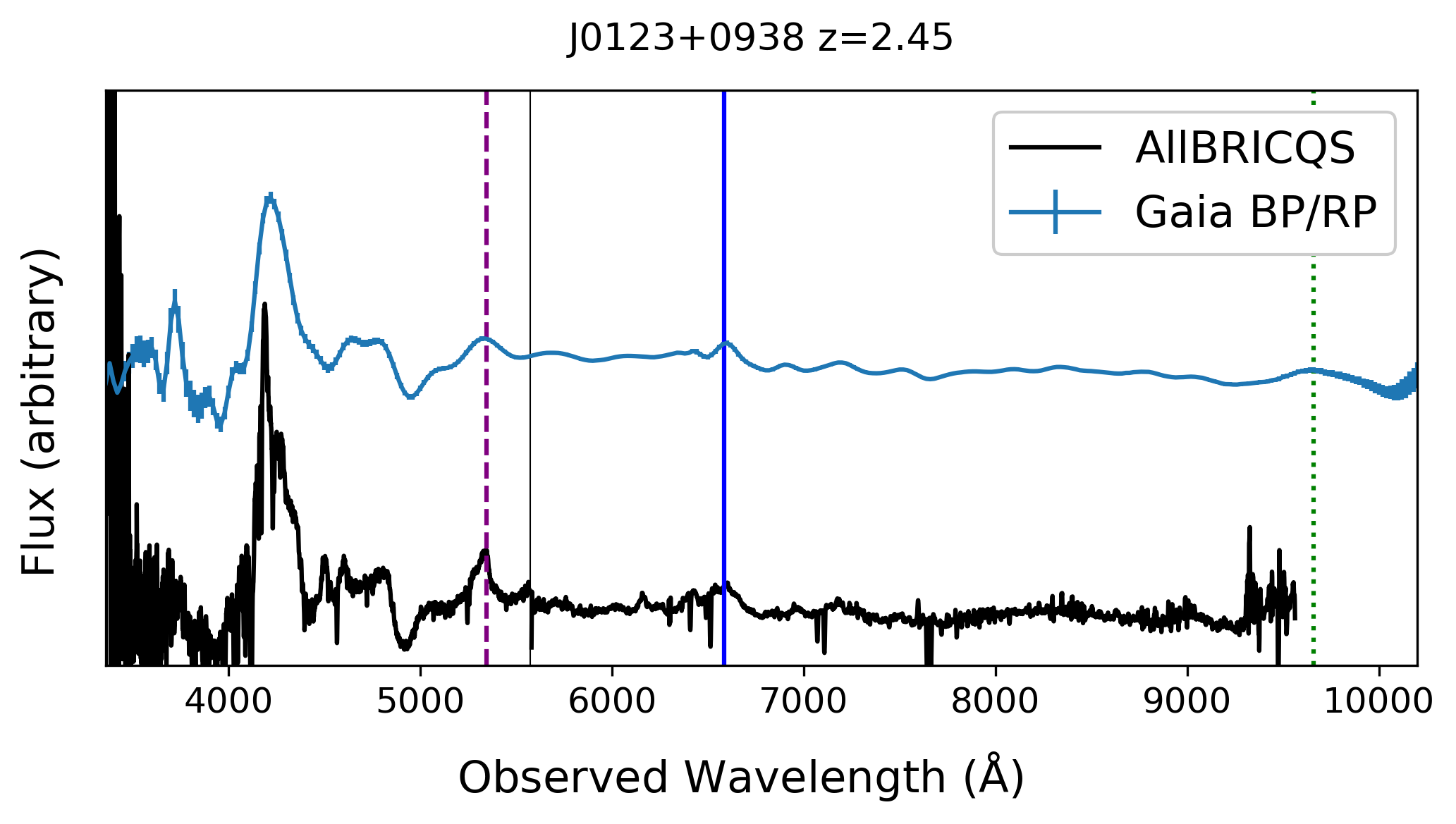}
\caption{Flux versus wavelength for the $R\sim3000$ spectrum of AllBRICQS (black line) and the {\it Gaia} low-resolution $B_{P}/R_{P}$ spectral reconstruction (blue points with errorbars) for four example quasars. Top labels indicate the AllBRICQS redshift as well as the {\it Gaia} classifications (contained in GLEAN or other variability samples, and/or a DSC-labelled quasar). Vertical lines indicate the positions of various quasar emissions lines: H$\alpha$ (red, dashed), H$\beta$ (orange, dot-dashed), \ion{Mg}{ii} (green, dotted), \ion{C}{iii}] (blue, solid), \ion{C}{iv} (purple, dashed). The thin, black vertical line at 5575~\AA\ indicates where the blue and red arms of the WiFeS spectrograph are spliced together. The {\it Gaia} spectra broaden and blend spectral features, which complicates the efforts to classify the sources and determine redshifts, let alone any more detailed analysis.}
\label{fig:spec_compare}
\end{figure*}

The challenges inherent in classifying quasars from {\it Gaia}'s low-resolution $B_{P}/R_{P}$ spectroscopy are evident when comparing against our $R\sim3000$ spectra from WiFeS (Figure~\ref{fig:spec_compare}). The blending of nearby emission lines, which can exhibit an intrinsic dispersion in relative fluxes, shifts the peak of the line emission, and the "ringing" in the spectra reconstructed from the DR3 basis functions, both at the blue end and in regions around strong emission lines, often makes an assessment of the true underlying spectrum difficult. In terms of further analysis of the spectra, studies of absorption features (both broad and narrow) and emission line widths (e.g., for BH mass estimation) are only possible with higher resolving power than provided by the $B_{P}/R_{P}$ spectra.

\subsection{{\it Gaia}-based redshift estimates}

Finally, we consider the performance of {\it Gaia}'s quasar classifier \cite[QSOC;][]{2022arXiv220606710D} on the AllBRICQS objects. QSOC further analyses the DSC-labelled quasars (including those with quasar probabilities as low as 1\%) to estimate redshifts from the $B_{P}/R_{P}$ spectra. The method, trained on SDSS quasars, produced redshifts for 130 of the AllBRICQS quasars, including 96 quasars for which flags\_qsoc = 0 (indicating reliable estimates). Among the latter, we find there to be one outlier, J2059-1632, a $z=0.8$ quasar for which QSOC estimated $z=3.146682\pm0.0089$ (mistaking \ion{Mg}{ii} for Ly$\alpha$). Amongst the remaining 95 quasars, the QSOC redshift estimate is 0.001 too low in the mean, with a root-mean-square variation of less than 0.005. With a typical QSOC redshift uncertainty of 0.001, the differences are likely dominated by the limited redshift precision to which AllBRICQS aspires, and thus we confirm that QSOC is a highly effective tool for estimating quasar redshifts.

\section{Discussion}\label{sec:discussion}

The sources discovered in AllBRICQS are those which have eluded detection in the first 50+ years of quasar searches, but many had already come under suspicion from other selection methods. In Section~\ref{sec:Gaia_comparison}, we described the high degree of completeness for the {\it Gaia} quasar classification, which is remarkable given their exclusive reliance on {\it Gaia}'s optical dataset (astrometric, photometric, and low-resolution spectroscopic). Yet the existence of AllBRICQS quasars that escaped the {\it Gaia} criteria confirms the value of multi-wavelength approaches. 

In addition to the {\it Gaia} variability and spectroscopic classifications, Table~\ref{tab:sample} indicates the objects which were flagged as Primary or Secondary quasar candidates from the Ultraviolet Quasar Survey \cite[UVQS;][]{2016AJ....152...25M}, which combined photometry from the {\it Galaxy Evolution Explorer} \cite[{\it GALEX};][]{2005ApJ...619L...1M} and {\it WISE} satellites. Apart from Galactic extinction, these UV/optical probes should provide the most complete samples of unobscured quasars at low to moderate redshift.

At higher energies, we anticipate the majority of the real quasars amongst the AllBRICQS candidates to be detected by {\it SRG}/eROSITA \cite[the {\it Spectrum-Roentgen-Gamma} satellite's extended ROentgen Survey with an Imaging Telescope Array;][]{2021A&A...647A...1P}. In addition, a small number of AllBRICQS quasars are also known to be associated with quasar-like radio emission, but did not have a spectroscopically confirmed redshift.

Outliers from the AllBRICQS selection criteria (in either {\it Gaia} astrometry or {\it WISE} colours) could still be identified at other wavelengths, and many of those bright quasars may be observed as part of the Black Hole Mapper\footnote{See \url{https://www.sdss5.org/mappers/black-hole-mapper/}.} (BHM) within SDSS-V \citep{2017arXiv171103234K}. Moreover, the BHM program will greatly expand the samples at fainter optical magnitudes, particularly as SDSS extends its optical spectroscopy footprint to the Southern hemisphere.

Early large-area surveys for blue objects were limited to high Galactic latitude, with the Cal\'{a}n-Tololo Survey \citep{1988ASPC....1..410M,1996RMxAA..32...35M} and Hamburg Quasar Survey \citep{1995A&AS..111..195H,1999A&AS..134..483H} extending closest to the Galactic Plane ($|b|>20^{\circ}$) in the Southern and Northern hemispheres, respectively. Even DR1 of the UVQS, which focused on FUV-detected sources from {\it GALEX}, was incomplete in its coverage near the Plane because of the failure of the FUV detector prior to the relaxing of the satellite's initial safety constraints on overall count rates \cite[see][]{2014Ap&SS.354..103B}. 

AllBRICQS is covering the sky down to $|b|>10^{\circ}$, but is still constrained from fully exploring behind the Galactic Plane because of source confusion arising from the limited spatial resolution of {\it WISE} ($\sim6$~arcsec in $W1$ and $W2$). Dedicated quasar searches within the Galactic Plane require alternative selection criteria to mitigate the effects of dust extinction, but have begun yielding significant samples \citep{2007ApJ...664...64I,2021ApJS..254....6F,2022ApJS..261...32F}. Beyond photometric searches, additional low-latitude quasars may be uncovered thanks to the near-IR resolving power of the forthcoming {\it SPHEREx} satellite's all-sky spectrophotometric survey\footnote{See \url{https://spherex.caltech.edu}.} \citep{2020SPIE11443E..0IC}.

AllBRICQS will complete observations of the target sample at declination $\delta<+20^\circ$ with the ANU 2.3m telescope at SSO. The intention for the Northern targets is to collect spectra at the Bohyunsan Optical Astronomy Observatory (BOAO) 1.8-meter telescope and the Seoul National University Astronomical Observatory (SAO) 1.0-meter telescope in South Korea, and the Yunnan Observatories (YNAO) Lijiang 2.4-meter telescope at the Gaomeigu site near Lijiang, China. These observations will confirm classifications and determine redshifts, as well as enabling virial estimates of BH masses \cite[see][and references therein]{2021iSci...24j2557C}. They will also reveal unusual optical spectral properties and foreground absorbers, which may flag specific targets for detailed follow-up at other facilities.

In particular, the two bright quasars at $z>2.5$ --- J0431-0838 at $z=2.53$ with SMSS DR3 $i=16.42$~ABmag, and J0529-4351 at $z=3.93$ with $i=16.07$~ABmag --- may be appealing targets for the Sandage Test \citep{1962ApJ...136..319S}. With high-stability wavelength calibration and extremely large telescopes, it should soon be possible to directly detect the expansion of the Universe through the wavelength drift of intervening absorbers over time baselines of decades \citep{2008MNRAS.386.1192L}. These two AllBRICQS quasars, and subsequent discoveries at similar redshifts, would supplement the samples recently compiled by QUBRICS \cite[QUasars as BRIght beacons for Cosmology in the Southern Hemisphere;][]{2019ApJ...887..268C,2020ApJS..250...26B} and ELQS \cite[Extremely Luminous Quasar Survey;][]{2017ApJ...851...13S,2019ApJ...871..258S}.

Finally, we note that a few AllBRICQS quasars were catalogued as AGNs by the Edinburgh-Cape (EC) Blue Object Survey, but because of the EC Survey's focus on hot, evolved stars in the Milky Way, the quasar spectra were not presented and no redshifts were recorded. We expect that many of the AllBRICQS candidates have been observed by groups looking for objects other than quasars, who seemingly had little incentive to publish those spectra --- a situation which nearly held true for the case of J1144-4308 \citep{2022arXiv220604204O}. Yet even if AllBRICQS presents spectra of those objects in future papers, the historical datasets could hold enormous value for studies of variability properties and searches for changing-look AGNs \cite[e.g.,][]{2015ApJ...800..144L}, and we encourage readers to dust off their old datasets and publish their quasar spectra.

\begin{acknowledgement}
For contributions to the observing, we thank Thomas Nordlander, and we thank Cassidy Mihalenko and Jemma Pilossof, as well as their supervisor, Katie Auchettl. We thank the anonymous referee and the editorial staff at PASA for suggestions that improved the manuscript. We thank the RSAA Software Group and the staff at SSO for their dedicated efforts to automate the ANU 2.3m telescope. We thank David Kilkenny for searching through the records of the Edinburgh-Cape spectra. We thank Scott Anderson for helpful discussions about SDSS-V's Black Hole Mapper. We acknowledge the traditional owners of the land on which the telescopes of Siding Spring Observatory stand, the Kamilaroi people, and pay our respects to their elders, past and present.

CAO was supported by the Australian Research Council (ARC) through Discovery Project DP190100252. SL is grateful to the Research School of Astronomy \& Astrophysics at Australian National University for funding his Ph.D. studentship.

This work has made use of data from the European Space Agency (ESA) mission
{\it Gaia} (\url{https://www.cosmos.esa.int/gaia}), processed by the {\it Gaia}
Data Processing and Analysis Consortium (DPAC,
\url{https://www.cosmos.esa.int/web/gaia/dpac/consortium}). Funding for the DPAC
has been provided by national institutions, in particular the institutions
participating in the {\it Gaia} Multilateral Agreement. This work has made use of the {\sc Python} package \texttt{GaiaXPy}, developed and maintained by members of the Gaia Data Processing and Analysis Consortium (DPAC), and in particular, Coordination Unit 5 (CU5), and the Data Processing Centre located at the Institute of Astronomy, Cambridge, UK (DPCI).

This publication makes use of data products from the Wide-field Infrared Survey Explorer, which is a joint project of the University of California, Los Angeles, and the Jet Propulsion Laboratory/California Institute of Technology, and NEOWISE, which is a project of the Jet Propulsion Laboratory/California Institute of Technology. WISE and NEOWISE are funded by the National Aeronautics and Space Administration.

This research uses services or data provided by the Astro Data Lab at National Science Foudation's National Optical-Infrared Astronomy Research Laboratory. NOIRLab is operated by the Association of Universities for Research in Astronomy (AURA), Inc. under a cooperative agreement with the NSF.

This project used data obtained with the Dark Energy Camera (DECam), which was constructed by the Dark Energy Survey (DES) collaboration. Funding for the DES Projects has been provided by the US Department of Energy, the US National Science Foundation, the Ministry of Science and Education of Spain, the Science and Technology Facilities Council of the United Kingdom, the Higher Education Funding Council for England, the National Center for Supercomputing Applications at the University of Illinois at Urbana-Champaign, the Kavli Institute for Cosmological Physics at the University of Chicago, Center for Cosmology and Astro-Particle Physics at the Ohio State University, the Mitchell Institute for Fundamental Physics and Astronomy at Texas A\&M University, Financiadora de Estudos e Projetos, Funda\c{c}\~{a}o Carlos Chagas Filho de Amparo \`{a} Pesquisa do Estado do Rio de Janeiro, Conselho Nacional de Desenvolvimento Cient\'{i}fico e Tecnol\'{o}gico and the Minist\/{e}rio da Ci\^{e}ncia, Tecnologia e Inova\c{c}\~{a}o, the Deutsche Forschungsgemeinschaft and the Collaborating Institutions in the Dark Energy Survey.

The Collaborating Institutions are Argonne National Laboratory, the University of California at Santa Cruz, the University of Cambridge, Centro de Investigaciones En\/{e}rgeticas, Medioambientales y Tecnol\'{o}gicas–Madrid, the University of Chicago, University College London, the DES-Brazil Consortium, the University of Edinburgh, the Eidgen\"{o}ssische Technische Hochschule (ETH) Z\"{u}rich, Fermi National Accelerator Laboratory, the University of Illinois at Urbana-Champaign, the Institut de Ci\`{e}ncies de l’Espai (IEEC/CSIC), the Institut de F\'{i}sica d’Altes Energies, Lawrence Berkeley National Laboratory, the Ludwig-Maximilians Universit\"{a}t M\"{u}nchen and the associated Excellence Cluster Universe, the University of Michigan, NSF’s NOIRLab, the University of Nottingham, the Ohio State University, the OzDES Membership Consortium, the University of Pennsylvania, the University of Portsmouth, SLAC National Accelerator Laboratory, Stanford University, the University of Sussex, and Texas A\&M University.

Based on observations at Cerro Tololo Inter-American Observatory, a program of NOIRLab (NOIRLab Prop. ID 2014B-0193; PI: F.~Walter), which is managed by the AURA under a cooperative agreement with the NSF.

The Pan-STARRS1 Surveys (PS1) and the PS1 public science archive have been made possible through contributions by the Institute for Astronomy, the University of Hawaii, the Pan-STARRS Project Office, the Max-Planck Society and its participating institutes, the Max Planck Institute for Astronomy, Heidelberg and the Max Planck Institute for Extraterrestrial Physics, Garching, The Johns Hopkins University, Durham University, the University of Edinburgh, the Queen's University Belfast, the Harvard-Smithsonian Center for Astrophysics, the Las Cumbres Observatory Global Telescope Network Incorporated, the National Central University of Taiwan, the Space Telescope Science Institute, the National Aeronautics and Space Administration under Grant No. NNX08AR22G issued through the Planetary Science Division of the NASA Science Mission Directorate, the National Science Foundation Grant No. AST-1238877, the University of Maryland, Eotvos Lorand University (ELTE), the Los Alamos National Laboratory, and the Gordon and Betty Moore Foundation.

The national facility capability for SkyMapper has been funded through ARC LIEF grant LE130100104 from the Australian Research Council, awarded to the University of Sydney, the Australian National University, Swinburne University of Technology, the University of Queensland, the University of Western Australia, the University of Melbourne, Curtin University of Technology, Monash University and the Australian Astronomical Observatory. SkyMapper is owned and operated by The Australian National University's Research School of Astronomy and Astrophysics. The survey data were processed and provided by the SkyMapper Team at ANU. The SkyMapper node of the All-Sky Virtual Observatory (ASVO) is hosted at the National Computational Infrastructure (NCI). Development and support of the SkyMapper node of the ASVO has been funded in part by Astronomy Australia Limited (AAL) and the Australian Government through the Commonwealth's Education Investment Fund (EIF) and National Collaborative Research Infrastructure Strategy (NCRIS), particularly the National eResearch Collaboration Tools and Resources (NeCTAR) and the Australian National Data Service Projects (ANDS).

\end{acknowledgement}

\bibliography{bib}

\appendix

\section{AllBRICQS Confirmed Quasars}\label{app:sample}

We list the properties of the confirmed AllBRICQS quasars in Table~\ref{tab:sample}, arranged in order of ascending RA. The $W1-W2$ colour is taken from the more accurate of CatWISE2020 \citep{2021ApJS..253....8M} and AllWISE\footnote{See \url{https://wise2.ipac.caltech.edu/docs/release/allwise/expsup/index.html}.}. The "UVQS" column indicates whether the source was a primary ("1") or secondary ("2") candidate in the FUV-selected sample from UVQS DR1 \citep{2016AJ....152...25M}. The "{\it Gaia}" column is a two-element code indicating (1) whether the source is included in the {\it Gaia} DR3 variabLE AgN sample \cite["G";][]{2022arXiv220706849C}, or only in another {\it Gaia} DR3 variability table\footnote{See \url{https://gea.esac.esa.int/archive/documentation/GDR3/Data_analysis/chap_cu7var/sec_cu7var_intro/ssec_cu7var_dataproducts.html}.} ("v"); and (2) whether the source has classlabel\_dsc='quasar' according to the {\it Gaia} Discrete Source Classifier ("D"). Sources not meeting one criteria or the other (or both) are indicated with "{--}".

In the table notes, we indicate two sources of known redshift for which optical spectra had not been previously published; 13 sources identified as quasars from the Edinburgh-Cape Blue Object Survey \citep{1997MNRAS.287..848S,2016MNRAS.459.4343K}, but without published redshifts (intentionally retained in the AllBRICQS target list); one source at $b=+19.3^{\circ}$ discovered by \citet{2022ApJS..261...32F} during preparation of this manuscript; one source with a low-quality redshift estimate from UVQS DR1, which we revise; and two quasars mistakenly catalogued as stars in the Lick Northern Proper Motion survey \cite[][with one also in Milliquas as a star with radio detection]{1987AJ.....94..501K}. We also flag four quasars taken from a magnitude bin 0.5~mag fainter than the nominal AllBRICQS limits.

\renewcommand\thetable{A.\arabic{table}}    
\begin{table*}[ht]
\begin{threeparttable}
\caption{Confirmed AllBRICQS Quasars}\label{tab:sample}
\begin{tabular}{llrrccclccl}
\toprule
Name & {\it Gaia} DR3 {\sc source\_id} & RA (J2000) & Dec (J2000) & $B_{P}$ & $R_{P}$ & $W1-W2$ & Redshift & UVQS & {\it Gaia} & Notes \\
 &  & deg & deg & mag & mag & mag & & & & \\
\midrule
J0000-7524 & 4685240533123833088 & 0.1563 & -75.4118 & 16.46 & 15.75 & 1.05 & 0.23 & 1 & GD & \\ % 
J0006-6457 & 4900255628276737792 & 1.5663 & -64.9616 & 16.51 & 15.81 & 1.19 & 1.724 & {--} & GD & \\ % 
J0010-6959 & 4702951122826363904 & 2.6084 & -69.9870 & 15.64 & 15.12 & 1.19 & 0.423 & {--} & GD & \\ % 
J0010-0702 & 2441707363251993216 & 2.6482 & -7.0423 & 16.48 & 16.05 & 0.99 & 0.207 & 2 & GD & \\ % 
J0014-2235 & 2361107938255080832 & 3.6834 & -22.5896 & 16.39 & 15.85 & 1.40 & 1.44 & 1 & GD & \\ % 
J0028-4054 & 4993849326503326720 & 7.2018 & -40.9038 & 16.20 & 15.71 & 1.04 & 0.219 & {--} & GD & \\ % 
J0035-7820 & 4635525182864394368 & 8.9961 & -78.3403 & 16.43 & 15.84 & 1.32 & 0.96 & {--} & GD & 1 \\ % 
J0056+1141 & 2582790132617218048 & 14.2319 & 11.6984 & 17.32 & 15.99 & 0.98 & 1.65 & {--} & {--}{--} & 2 \\ % 
J0117-1712 & 2355371511215136384 & 19.4378 & -17.2117 & 16.39 & 15.83 & 0.76 & 0.68 & {--} & GD & \\ % 
J0123+0938 & 2579205655991234048 & 20.8272 & 9.6372 & 16.70 & 15.73 & 0.68 & 2.45 & {--} & G{--} & \\ % 
J0140-0653 & 2479251287293182592 & 25.1431 & -6.8958 & 16.13 & 15.50 & 0.98 & 0.2 & 2 & GD & \\ % 
J0141-1607 & 2452322735700697600 & 25.3660 & -16.1270 & 16.79 & 15.93 & 1.18 & 0.66 & {--} & GD & \\ % 
J0146-2608 & 5025447435259941376 & 26.6884 & -26.1460 & 16.58 & 15.92 & 1.12 & 0.614 & {--} & {--}D & \\ % 
J0154-5559 & 4719544299477963520 & 28.6425 & -55.9846 & 15.57 & 14.94 & 1.17 & 1.6 & {--} & GD & \\ % 
J0159-3205 & 5018050883101214848 & 29.8456 & -32.0894 & 16.76 & 16.00 & 1.48 & 1.056 & {--} & vD & \\ % 
J0201+1134 & 2574519674872605184 & 30.4064 & 11.5759 & 16.58 & 15.83 & 1.01 & 0.203 & 1 & GD & \\ % 
J0207-2354 & 5121675902648838656 & 31.9490 & -23.9029 & 14.90 & 14.29 & 0.92 & 0.076 & {--} & GD & 3 \\ %
J0210-5321 & 4743906728370668288 & 32.6818 & -53.3561 & 16.78 & 15.95 & 1.41 & 1.52 & {--} & v{--} & \\ % 
J0219+1925 & 87024356769335808 & 34.9021 & 19.4266 & 16.27 & 15.50 & 1.19 & 0.824 & {--} & G{--} & \\ % 
J0220-2519 & 5119524364551391872 & 35.2302 & -25.3242 & 16.54 & 15.93 & 1.24 & 2.173 & {--} & GD & \\ % 
J0241-1719 & 5132542547864397056 & 40.4892 & -17.3173 & 15.98 & 15.51 & 1.00 & 0.187 & 2 & GD & \\ % 
J0245-8035 & 4619847113421226240 & 41.4709 & -80.5927 & 16.19 & 15.70 & 1.32 & 0.927 & 1 & GD & \\ % 
J0252-2650 & 5072673246379455104 & 43.1902 & -26.8473 & 16.44 & 15.87 & 1.17 & 0.83 & {--} & GD & \\ % 
J0311-4655 & 4750861929689855360 & 47.9983 & -46.9300 & 16.90 & 15.67 & 1.12 & 0.5546 & {--} & G{--} & \\ % 
J0315-7434 & 4639775757378274944 & 48.9038 & -74.5695 & 16.41 & 15.96 & 1.23 & 0.691 & {--} & GD & \\ % 
J0316-0919 & 5166622494882955520 & 49.2104 & -9.3253 & 16.42 & 15.67 & 1.42 & 0.287 & {--} & GD & \\ % 
J0327-7224 & 4642275673158740224 & 51.7594 & -72.4010 & 16.06 & 15.41 & 0.86 & 0.126 & 2 & GD & 1 \\ % 
J0328-6225 & 4674156528202556160 & 52.0129 & -62.4254 & 16.15 & 15.56 & 1.06 & 0.18 & {--} & GD & 1 \\ % 
J0343-1711 & 5108385074813114496 & 55.8885 & -17.1857 & 15.64 & 15.07 & 1.28 & 0.66 & {--} & GD & 1 \\ % 
J0400-2257 & 5090217427573688192 & 60.1855 & -22.9534 & 16.33 & 15.58 & 0.95 & 0.163 & 2 & GD & \\ % 
J0405-2410 & 5089142453095475840 & 61.4342 & -24.1808 & 16.45 & 15.94 & 1.46 & 1.26 & {--} & GD & \\ % 
J0425-4410 & 4839258610114187264 & 66.3719 & -44.1693 & 15.26 & 14.85 & 1.09 & 0.39 & {--} & {--}D & 1 \\ %
J0427-1412 & 3175855537725581952 & 66.7904 & -14.2024 & 16.59 & 15.98 & 1.12 & 0.388 & {--} & GD & 1 \\ % 
J0431-0838 & 3197674349547449856 & 67.7730 & -8.6407 & 17.07 & 15.98 & 0.96 & 2.53 & {--} & G{--} & \\ % 
J0433-0641 & 3198628966157590784 & 68.3834 & -6.6968 & 16.49 & 16.20 & 1.28 & 0.81 & {--} & GD & \\ % 
\bottomrule
\multicolumn{2}{l}{Position information from Gaia DR3.}\\
\multicolumn{11}{l}{1. Known quasar from Edinburgh-Cape Blue Object Survey, but redshift not recorded.}\\
\multicolumn{11}{l}{2. Known quasar with $z=1.656$ 
from \citet{2016ApJS..224...11G}, but no spectrum presented. Data available from CfA Optical/Infrared Science Archive.}\\
\multicolumn{11}{l}{3. Known quasar with $z=0.076$ from \citet{2020ApJ...898...61I}, but no spectrum presented.}\\
\end{tabular}
\end{threeparttable}
\end{table*}

\setcounter{table}{0}
\begin{table*}[ht]
\begin{threeparttable}
\caption{Confirmed AllBRICQS Quasars (cont.)}
\begin{tabular}{llrrccclccl}
\toprule
Name & {\it Gaia} DR3 {\sc source\_id} & RA (J2000) & Dec (J2000) & $B_{P}$ & $R_{P}$ & $W1-W2$ & Redshift & UVQS & {\it Gaia} & Notes \\
 &  & deg & deg & mag & mag & mag & & & & \\
\midrule
J0500-6322 & 4665166852415732352 & 75.0027 & -63.3771 & 15.96 & 15.52 & 1.23 & 1.18 & {--} & GD & \\ % 
J0502-2002 & 2975142095258671744 & 75.5014 & -20.0368 & 16.24 & 15.64 & 1.16 & 0.375 & {--} & GD & \\ % 
J0502-6227 & 4664566278549403392 & 75.7110 & -62.4613 & 16.24 & 15.80 & 1.11 & 0.382 & 2 & GD & 1 \\ % 
J0504+0055 & 3228636356466440320 & 76.2366 & 0.9272 & 15.90 & 15.06 & 0.88 & 2.34 & {--} & G{--} & \\ % 
J0513-3320 & 4826131231553088896 & 78.3498 & -33.3358 & 16.53 & 15.81 & 1.16 & 0.419 & {--} & GD & \\ % 
J0514-1618 & 2983155095483011584 & 78.6401 & -16.3106 & 16.54 & 15.87 & 1.23 & 0.2272 & {--} & GD & \\ % 
J0529-4351 & 4805630493655815040 & 82.3159 & -43.8645 & 16.91 & 15.69 & 0.38 & 3.93 & {--} & v{--} & \\ % 
J0530+0042 & 3221158234288319744 & 82.6800 & 0.7118 & 16.28 & 15.70 & 1.15 & 0.7 & {--} & GD & \\ % 
J0533-1434 & 2984416785075086336 & 83.2543 & -14.5704 & 16.30 & 15.60 & 1.00 & 0.139 & 2 & GD & \\ % 
J0546-4630 & 4795962449256424448 & 86.6186 & -46.5021 & 16.43 & 15.90 & 1.08 & 0.279 & 2 & GD & \\ % 
J0555-5100 & 4792646051605064832 & 88.8761 & -51.0048 & 16.43 & 15.88 & 1.02 & 0.263 & {--} & vD & \\ % 
J0600-4105 & 2882337747595526528 & 90.0612 & -41.0911 & 16.04 & 15.48 & 1.40 & 1.19 & {--} & GD & \\ % 
J0603-3110 & 2890998291450368896 & 90.8093 & -31.1790 & 16.37 & 15.94 & 1.15 & 0.726 & {--} & GD & \\ % 
J0624-6324 & 5477304692317531392 & 96.1551 & -63.4118 & 15.94 & 15.38 & 1.19 & 0.576 & {--} & GD & \\ % 
J0624-2545 & 2900007414852025088 & 96.2033 & -25.7659 & 15.16 & 14.46 & 0.83 & 0.863 & {--} & G{--} & \\ % 
J0634-6945 & 5278908676764365824 & 98.7319 & -69.7589 & 15.99 & 15.36 & 1.34 & 1.37 & {--} & {--}{--} & \\ % 
J0639-2653 & 2919897167882522496 & 99.9759 & -26.8965 & 16.26 & 15.55 & 1.15 & 0.705 & {--} & G{--} & \\ % 
J0712-5659 & 5486909991537522688 & 108.0864 & -56.9940 & 15.49 & 15.00 & 1.36 & 0.907 & {--} & GD & \\ % 
J0715-4951 & 5505302725127883136 & 108.7923 & -49.8524 & 15.04 & 14.05 & 0.80 & 0.1156 & {--} & v{--} & \\ % 
J0749+0203 & 3088492604393071488 & 117.4535 & 2.0600 & 15.90 & 15.44 & 1.13 & 0.374 & {--} & GD & \\ % 
J0818-7933 & 5208653628959060608 & 124.5859 & -79.5603 & 15.95 & 15.34 & 1.43 & 1.29 & {--} & GD & \\ % 
J0833-0628 & 5754948798717254144 & 128.4389 & -6.4800 & 16.25 & 15.77 & 1.36 & 1.2 & {--} & GD & 4 \\ % 
J0835-0833 & 5753659140297301760 & 128.8037 & -8.5636 & 15.99 & 15.63 & 1.08 & 0.585 & {--} & GD & \\ % 
J0854-0718 & 5756861914589220736 & 133.6467 & -7.3104 & 16.24 & 15.53 & 1.03 & 1.14 & {--} & GD & \\ % 
J0907-2000 & 5680217153745527936 & 136.9000 & -20.0005 & 16.32 & 15.83 & 1.16 & 0.57 & 1 & GD & 5 \\ % 
J0930-7528 & 5215913601159378944 & 142.6517 & -75.4817 & 16.14 & 15.50 & 1.10 & 0.35 & {--} & GD & \\ % 
J0934-3325 & 5438016461800707712 & 143.6061 & -33.4233 & 16.39 & 15.73 & 1.12 & 0.86 & {--} & {--}{--} & \\ % 
J1037-2223 & 5474848486419684096 & 159.4465 & -22.3887 & 16.13 & 15.70 & 1.30 & 0.96 & {--} & GD & 1 \\ % 
J1040-3324 & 5444798279580321536 & 160.1045 & -33.4127 & 16.16 & 15.55 & 1.21 & 1.04 & {--} & GD & \\ % 
J1048-3401 & 5449737908584117376 & 162.1377 & -34.0227 & 16.01 & 15.44 & 1.21 & 1.19 & {--} & G{--} & \\ % 
J1049-3001 & 5454621252040188672 & 162.4913 & -30.0177 & 16.50 & 15.98 & 1.14 & 1.44 & {--} & GD & \\ % 
J1120-2939 & 3482845025356707584 & 170.2026 & -29.6608 & 16.49 & 16.04 & 1.17 & 0.65 & 1 & GD & 1 \\ % 
J1128-7435 & 5225252406250984448 & 172.0981 & -74.5935 & 16.36 & 15.21 & 1.28 & 1.48 & {--} & {--}{--} & \\ % 
J1205+0845 & 3905774372003183360 & 181.3565 & 8.7508 & 16.36 & 15.81 & 0.92 & 0.935 & {--} & GD & \\ % 
J1215-3221 & 3469391607236907392 & 183.8098 & -32.3505 & 16.50 & 15.85 & 0.93 & 0.1727 & 2 & vD & \\ % 
\bottomrule
\multicolumn{2}{l}{Position information from Gaia DR3.}\\
\multicolumn{11}{l}{1. Known quasar from Edinburgh-Cape Blue Object Survey, but redshift not recorded.}\\
\multicolumn{11}{l}{4. Known quasar with $z=1.201$ recently discovered by \citet{2022ApJS..261...32F}.}\\
\multicolumn{11}{l}{5. UVQS DR1 source catalogued as $z=1.279$ having z\_qual=0 ("no estimate possible").}
\end{tabular}
\end{threeparttable}
\end{table*}

\setcounter{table}{0}
\begin{table*}[ht]
\begin{threeparttable}
\caption{Confirmed AllBRICQS Quasars (cont.)}
\begin{tabular}{llrrccclccl}
\toprule
Name & {\it Gaia} DR3 {\sc source\_id} & RA (J2000) & Dec (J2000) & $B_{P}$ & $R_{P}$ & $W1-W2$ & Redshift & UVQS & {\it Gaia} & Notes \\
 &  & deg & deg & mag & mag & mag & & & & \\
\midrule
J1227-4133 & 6146727950157752320 & 186.9019 & -41.5519 & 16.54 & 15.86 & 1.29 & 0.99 & {--} & GD & \\ % 
J1249-3545 & 6155619047856315776 & 192.3628 & -35.7587 & 16.77 & 15.94 & 1.07 & 0.39 & {--} & {--}{--} & \\ % 
J1252-3928 & 6152529110943401216 & 193.1630 & -39.4697 & 16.59 & 15.69 & 0.91 & 0.2345 & {--} & GD & \\ % 
J1304-2318 & 3504140465345674624 & 196.2204 & -23.3147 & 15.40 & 14.60 & 1.10 & 0.3785 & {--} & GD & \\ % 
J1333-2249 & 6194977063004413312 & 203.3060 & -22.8319 & 16.70 & 15.55 & 1.12 & 0.8 & {--} & G{--} & \\ % 
J1345-4847 & 6094779942759884416 & 206.3290 & -48.7969 & 16.62 & 15.63 & 1.32 & 0.737 & {--} & G{--} & \\ % 
J1350-2924 & 6176765302157977856 & 207.6995 & -29.4011 & 16.54 & 15.97 & 1.32 & 1.09 & {--} & vD & \\ % 
J1357-3352 & 6170515643706741632 & 209.3862 & -33.8703 & 16.34 & 15.88 & 1.32 & 0.99 & {--} & GD & \\ % 
J1409-3704 & 6120955535042511744 & 212.4075 & -37.0801 & 16.39 & 15.77 & 1.38 & 1.05 & {--} & GD & \\ % 
J1410-3824 & 6117445893933200128 & 212.5764 & -38.4076 & 16.91 & 16.30 & 1.04 & 1.65 & {--} & {--}D & 6 \\ % 
J1416-1330 & 6302999235710077312 & 214.1154 & -13.5025 & 16.91 & 16.29 & 1.29 & 1.13 & {--} & GD & 6 \\ % 
J1419-7303 & 5797309644547609472 & 214.8014 & -73.0636 & 16.37 & 15.68 & 1.04 & 0.32 & {--} & G{--} & \\ % 
J1422-2453 & 6272600006945299072 & 215.5091 & -24.8835 & 16.75 & 15.83 & 1.25 & 1.35 & {--} & GD & \\ % 
J1424-3833 & 6116655241990813056 & 216.2170 & -38.5578 & 16.40 & 15.80 & 1.24 & 0.92 & {--} & vD & \\ % 
J1427-4251 & 6102649972113553024 & 216.8860 & -42.8549 & 15.49 & 14.79 & 0.92 & 0.25 & {--} & GD & \\ % 
J1455-4744 & 5905051652260501248 & 223.8729 & -47.7491 & 16.17 & 15.30 & 0.86 & 0.202 & {--} & GD & \\ % 
J1458-1621 & 6306813128014139776 & 224.7167 & -16.3514 & 16.95 & 15.89 & 1.23 & 0.587 & {--} & G{--} & \\ % 
J1459-7714 & 5791742855834137472 & 224.7994 & -77.2455 & 15.98 & 15.53 & 1.21 & 0.648 & {--} & GD & \\ % 
J1501-1053 & 6313702809608036736 & 225.2787 & -10.8903 & 16.71 & 16.12 & 1.13 & 0.72 & {--} & vD & 6 \\ % 
J1509-3950 & 6005386039652374784 & 227.3979 & -39.8405 & 16.98 & 16.50 & 1.32 & 0.91 & {--} & {--}D & 6 \\ % 
J1518-1736 & 6258932798238246912 & 229.5049 & -17.6045 & 16.28 & 15.36 & 1.13 & 0.68 & {--} & GD & \\ % 
J1518-2308 & 6251379947930795520 & 229.5082 & -23.1337 & 16.47 & 15.95 & 1.03 & 0.93 & {--} & GD & \\ % 
J1527-7828 & 5779460245798329984 & 231.7948 & -78.4740 & 16.03 & 15.46 & 1.25 & 0.95 & {--} & {--}D & \\ % 
J1538-4004 & 6002928699894118656 & 234.6418 & -40.0758 & 16.62 & 15.89 & 1.02 & 0.1518 & {--} & {--}D & \\ % 
J1544-2016 & 6241880304902964352 & 236.1084 & -20.2762 & 15.25 & 14.43 & 1.06 & 0.23 & {--} & GD & \\ % 
J1546-8422 & 5768119230033216128 & 236.6550 & -84.3732 & 16.57 & 15.97 & 1.15 & 0.374 & {--} & GD & \\ % 
J1547-1449 & 6263682211733086720 & 236.7656 & -14.8272 & 16.49 & 15.76 & 1.25 & 1.32 & {--} & GD & \\ % 
J1554-3209 & 6039417264563751808 & 238.6295 & -32.1637 & 15.69 & 14.79 & 1.17 & 0.278 & {--} & GD & \\ % 
J1559-6732 & 5822042647531326208 & 239.9957 & -67.5439 & 15.95 & 15.43 & 1.17 & 0.629 & {--} & GD & \\ % 
J1601-7202 & 5818662336458884224 & 240.4902 & -72.0431 & 15.66 & 14.87 & 0.96 & 0.1045 & {--} & GD & \\ % 
J1607-0740 & 4348755065537927296 & 241.8802 & -7.6732 & 16.76 & 15.78 & 0.94 & 0.2086 & {--} & GD & \\ % 
J1612-6958 & 5819359564272623104 & 243.0770 & -69.9812 & 16.08 & 15.47 & 1.21 & 1.37 & {--} & {--}D & \\ % 
J1618-3059 & 6037522256277200256 & 244.5586 & -30.9907 & 17.11 & 15.95 & 1.03 & 0.088 & {--} & {--}{--} & \\ % 
J1618-1424 & 4329648611456994176 & 244.6891 & -14.4070 & 16.67 & 15.71 & 0.98 & 0.207 & {--} & GD & \\ % 
J1619-7832 & 5778852799983279744 & 244.7576 & -78.5437 & 14.97 & 14.10 & 0.89 & 0.072 & 1 & v{--} & \\ % 
\bottomrule
\multicolumn{2}{l}{Position information from Gaia DR3.}\\
\multicolumn{11}{l}{6. Extended selection criteria: $B_{P}>16.5$ and $R_{P}>16$~mag.}\\
\end{tabular}
\end{threeparttable}
\end{table*}

\setcounter{table}{0}
\begin{table*}[ht]
\begin{threeparttable}
\caption{Confirmed AllBRICQS Quasars (cont.)}
\begin{tabular}{llrrccclccl}
\toprule
Name & {\it Gaia} DR3 {\sc source\_id} & RA (J2000) & Dec (J2000) & $B_{P}$ & $R_{P}$ & $W1-W2$ & Redshift & UVQS & {\it Gaia} & Notes \\
 &  & deg & deg & mag & mag & mag & & & & \\
\midrule
J1622+1400 & 4463914614888824064 & 245.5325 & 14.0136 & 16.38 & 15.83 & 0.92 & 0.93 & {--} & GD & \\ % 
J1654+0742 & 4442283510319657216 & 253.5757 & 7.7012 & 16.68 & 15.92 & 1.12 & 0.327 & {--} & GD & \\ % 
J1705+1354 & 4544604509076327936 & 256.4004 & 13.9051 & 16.74 & 15.79 & 1.45 & 1.3 & {--} & v{--} & \\ % 
J1720+1115 & 4540223539356310784 & 260.0047 & 11.2502 & 16.31 & 15.55 & 0.89 & 0.184 & 1 & GD & \\ % 
J1726+0128 & 4375126680125658880 & 261.5279 & 1.4796 & 16.68 & 15.75 & 1.49 & 1.6 & {--} & {--}{--} & \\ % 
J1728+1954 & 4554115250297917952 & 262.0982 & 19.9007 & 16.44 & 15.95 & 1.24 & 0.94 & {--} & GD & \\ % 
J1738+0042 & 4375310569150345216 & 264.7142 & 0.7106 & 16.65 & 15.81 & 1.08 & 0.212 & {--} & {--}D & \\ % 
J1817-4144 & 6724670225684293248 & 274.4713 & -41.7349 & 16.30 & 15.69 & 0.89 & 0.1978 & {--} & {--}D & \\ % 
J1904-1706 & 4088315532990245376 & 286.0108 & -17.1147 & 16.12 & 15.18 & 1.01 & 0.1995 & {--} & G{--} & \\ % 
J1904-5640 & 6642902913158050816 & 286.0246 & -56.6740 & 16.46 & 15.59 & 1.38 & 0.777 & {--} & G{--} & \\ % 
J1905-2639 & 6763933442298679040 & 286.3691 & -26.6542 & 15.99 & 14.97 & 1.07 & 0.121 & {--} & G{--} & \\ % 
J1910-4809 & 6661621033470406656 & 287.7193 & -48.1629 & 16.47 & 15.98 & 1.16 & 0.7677 & {--} & GD & \\ % 
J1916-1842 & 4084303552485486336 & 289.2401 & -18.7078 & 16.91 & 15.92 & 0.99 & 0.155 & {--} & {--}{--} & \\ % 
J1933-2129 & 6772236331332883840 & 293.3253 & -21.4843 & 16.20 & 15.66 & 1.11 & 0.855 & {--} & GD & \\ % 
J2000-4658 & 6671681324343864832 & 300.0483 & -46.9695 & 16.06 & 15.15 & 1.19 & 0.555 & {--} & G{--} & 1 \\ % 
J2006+1032 & 4300542923775269248 & 301.6039 & 10.5496 & 16.30 & 15.34 & 0.82 & 1.85 & {--} & G{--} & \\ % 
J2007+1235 & 1803066685791351168 & 301.9853 & 12.5882 & 16.20 & 15.62 & 1.12 & 0.4 & {--} & GD & \\ % 
J2034-0405 & 4218657791715405952 & 308.5555 & -4.0914 & 16.02 & 15.26 & 0.87 & 0.0995 & 2 & G{--} & \\ % 
J2048-6923 & 6376205952542312192 & 312.0243 & -69.3988 & 16.67 & 15.96 & 1.08 & 0.328 & 1 & GD & \\ % 
J2050-2530 & 6805397468883982464 & 312.5629 & -25.5098 & 16.55 & 15.95 & 1.25 & 0.883 & {--} & GD & \\ % 
J2055-4014 & 6773780084311817472 & 313.9932 & -40.2429 & 16.50 & 15.25 & 0.66 & 1.86 & {--} & {--}{--} & \\ % 
J2058-1452 & 6886716902195844352 & 314.5257 & -14.8761 & 15.86 & 15.27 & 0.92 & 0.135 & 1 & GD & \\ % 
J2059-1632 & 6883214987725624320 & 314.8057 & -16.5483 & 16.44 & 16.00 & 1.19 & 0.8 & {--} & GD & \\ % 
J2101+1350 & 1758570133100303872 & 315.2916 & 13.8339 & 16.97 & 15.93 & 0.88 & 1.419 & {--} & v{--} & \\ % 
J2101-3834 & 6774357224839696256 & 315.4377 & -38.5680 & 16.11 & 15.68 & 1.32 & 0.767 & {--} & GD & 1 \\ % 
J2102-7733 & 6368144814324065536 & 315.6306 & -77.5598 & 16.21 & 15.55 & 1.40 & 0.89 & 1 & G{--} & 1 \\ % 
J2107-6525 & 6449304925828127104 & 316.8300 & -65.4219 & 16.43 & 16.02 & 1.11 & 0.567 & 1 & GD & \\ % 
J2111-4949 & 6478751805723581952 & 317.9159 & -49.8229 & 16.52 & 15.96 & 0.92 & 0.336 & 1 & GD & \\ % 
J2112-4951 & 6478748678987384448 & 318.0231 & -49.8604 & 16.40 & 16.06 & 1.22 & 0.575 & 1 & GD & \\ % 
J2113-5840 & 6453491488148613760 & 318.4068 & -58.6791 & 15.87 & 15.44 & 0.85 & 0.99 & {--} & GD & \\ % 
J2114+1637 & 1784845196828559360 & 318.7141 & 16.6293 & 16.40 & 15.61 & 1.12 & 1.265 & {--} & v{--} & \\ % 
J2116-5931 & 6453194443914776320 & 319.0123 & -59.5296 & 16.62 & 15.93 & 1.05 & 1.75 & {--} & {--}D & \\ % 
J2119-0929 & 6895824775484040832 & 319.8336 & -9.4970 & 16.18 & 15.37 & 0.85 & 1.21 & {--} & G{--} & \\ % 
J2126-4529 & 6576106447897533440 & 321.7254 & -45.4994 & 16.99 & 15.78 & 1.11 & 0.167 & {--} & G{--} & \\ % 
J2128-7059 & 6372260702960045184 & 322.2374 & -70.9871 & 15.87 & 15.50 & 1.05 & 0.477 & {--} & GD & 1 \\ % 
\bottomrule
\multicolumn{2}{l}{Position information from Gaia DR3.}\\
\multicolumn{11}{l}{1. Known AGN from Edinburgh-Cape Blue Object Survey, but redshift not recorded.}\\
\end{tabular}
\end{threeparttable}
\end{table*}

\setcounter{table}{0}
\begin{table*}[ht]
\begin{threeparttable}
\caption{Confirmed AllBRICQS Quasars (cont.)}
\begin{tabular}{llrrccclccl}
\toprule
Name & {\it Gaia} DR3 {\sc source\_id} & RA (J2000) & Dec (J2000) & $B_{P}$ & $R_{P}$ & $W1-W2$ & Redshift & UVQS & {\it Gaia} & Notes \\
 &  & deg & deg & mag & mag & mag & & & & \\
\midrule
J2135+0858 & 1741107281406118656 & 323.7702 & 8.9668 & 16.43 & 15.94 & 0.24 & 0.25 & {--} & GD & \\ % 
J2149+1827 & 1774065898363122560 & 327.3104 & 18.4642 & 16.67 & 15.98 & 0.86 & 0.878 & {--} & G{--} & \\ % 
J2156-2400 & 6813438021321493120 & 329.2228 & -24.0157 & 16.51 & 15.42 & 1.31 & 0.87 & {--} & G{--} & \\ % 
J2207-1654 & 6826143908972604672 & 331.9871 & -16.9110 & 16.88 & 15.94 & 0.92 & 0.1288 & {--} & GD & \\ % 
J2234-6757 & 6385868911699682304 & 338.6292 & -67.9561 & 16.48 & 15.55 & 1.00 & 0.365 & {--} & G{--} & \\ % 
J2242-1316 & 2598238648944896640 & 340.7390 & -13.2830 & 15.93 & 15.37 & 1.05 & 0.274 & {--} & GD & 7 \\ % 
J2248-2446 & 6623320817223115776 & 342.1243 & -24.7827 & 16.65 & 15.94 & 1.38 & 1.62 & {--} & GD & \\ % 
J2306-2604 & 2382576796140530432 & 346.5987 & -26.0782 & 16.37 & 16.01 & 1.36 & 1.04 & {--} & GD & \\ % 
J2316-1941 & 2392891623957984768 & 349.1658 & -19.6871 & 16.49 & 16.08 & 1.25 & 0.99 & 1 & GD & \\ % 
J2320-0524 & 2633514757239529728 & 350.1815 & -5.4126 & 16.03 & 15.47 & 1.00 & 1.368 & {--} & GD & \\ % 
J2321-1521 & 2408315676151838080 & 350.3087 & -15.3584 & 16.44 & 15.99 & 1.18 & 0.665 & {--} & vD & \\ % 
J2324-4250 & 6535888953956911232 & 351.1461 & -42.8426 & 16.51 & 15.98 & 1.20 & 0.97 & {--} & GD & \\ % 
J2329-2133 & 2388531854195433344 & 352.4782 & -21.5658 & 15.81 & 15.28 & 1.12 & 0.32 & 2 & GD & 7 \\ % 
J2331-6642 & 6389695349603211008 & 352.8306 & -66.7006 & 16.50 & 15.73 & 0.96 & 0.274 & 1 & GD & \\ % 
J2343-4519 & 6531466172730996096 & 355.8729 & -45.3279 & 15.97 & 15.37 & 1.23 & 0.813 & {--} & GD & \\ % 
J2344-1121 & 2432988884582722560 & 356.0001 & -11.3665 & 16.47 & 16.10 & 1.25 & 0.65 & 1 & GD & \\ % 
\bottomrule
\multicolumn{2}{l}{Position information from Gaia DR3.}\\
\multicolumn{11}{l}{7. Reported as star in the Lick Northern Proper Motion survey \citep{1987AJ.....94..501K}.}\\
\end{tabular}
\end{threeparttable}
\end{table*}

\section{Gallery of AllBRICQS Quasar Spectra}\label{app:gallery}
\renewcommand\thefigure{A.\arabic{figure}}    
\setcounter{figure}{0}    

\begin{figure*}[t]
\centering
\includegraphics[width=\textwidth]{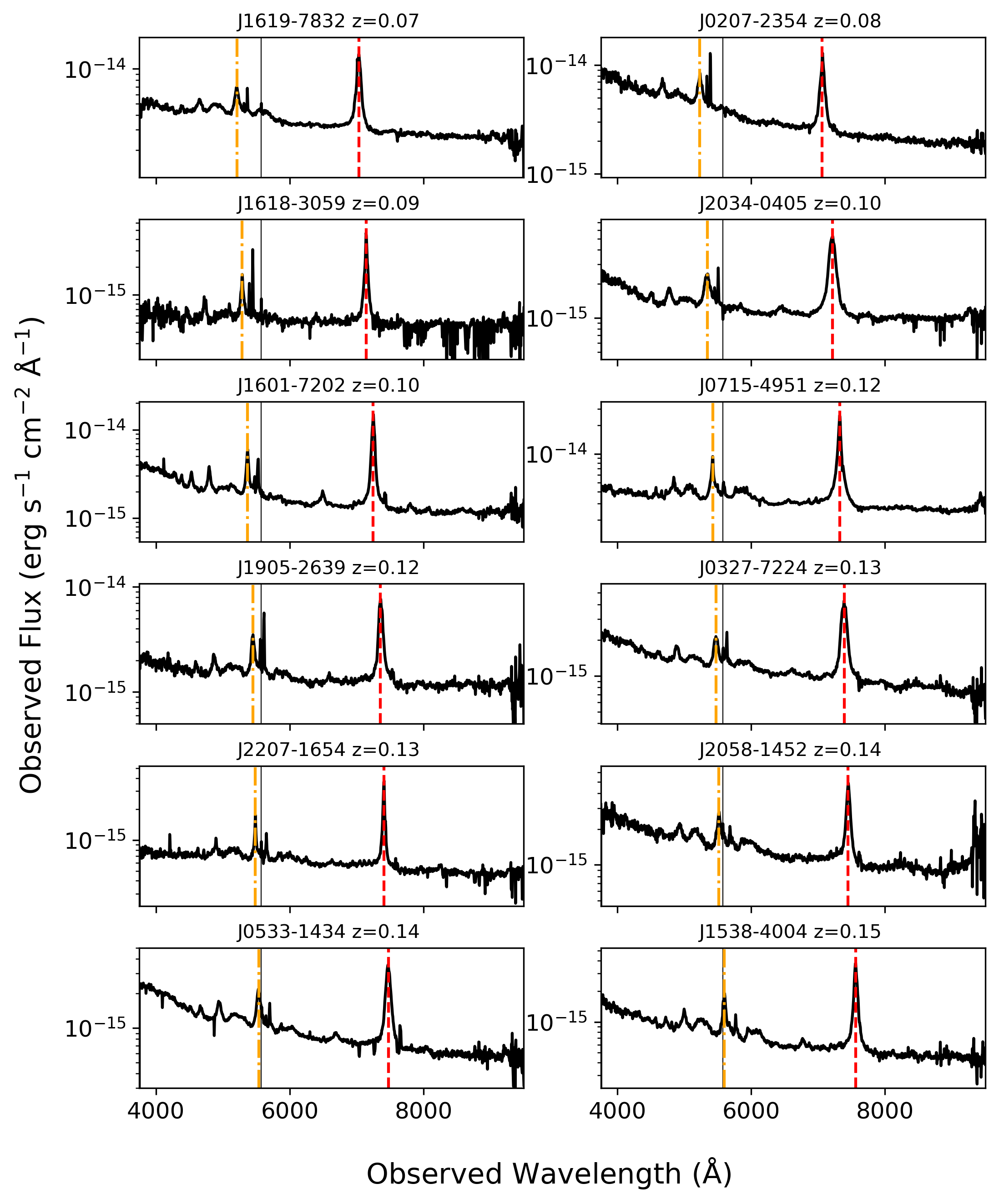}
\caption{WiFeS spectra of the AllBRICQS sample, shown with increasing redshift. The y-axis shows the flux on a logarithmic scale, normalised using Gaia DR3 photometry. Vertical lines indicate the positions of various quasar emissions lines: H$\alpha$ (red, dashed), H$\beta$ (orange, dot-dashed), \ion{Mg}{ii} (green, dotted), \ion{C}{iii}] (blue, solid), \ion{C}{iv} (purple, dashed). The thin, black vertical line at 5575~\AA\ indicates where the blue and red arms of the spectrograph are spliced together.}
\label{fig:gallery1}
\end{figure*}

We present the WiFeS spectra for the AllBRICQS quasars in Figures~\ref{fig:gallery1}-\ref{fig:galleryLast}, arranged in order of ascending redshift.

\begin{figure*}[t]
\centering
\includegraphics[width=\textwidth]{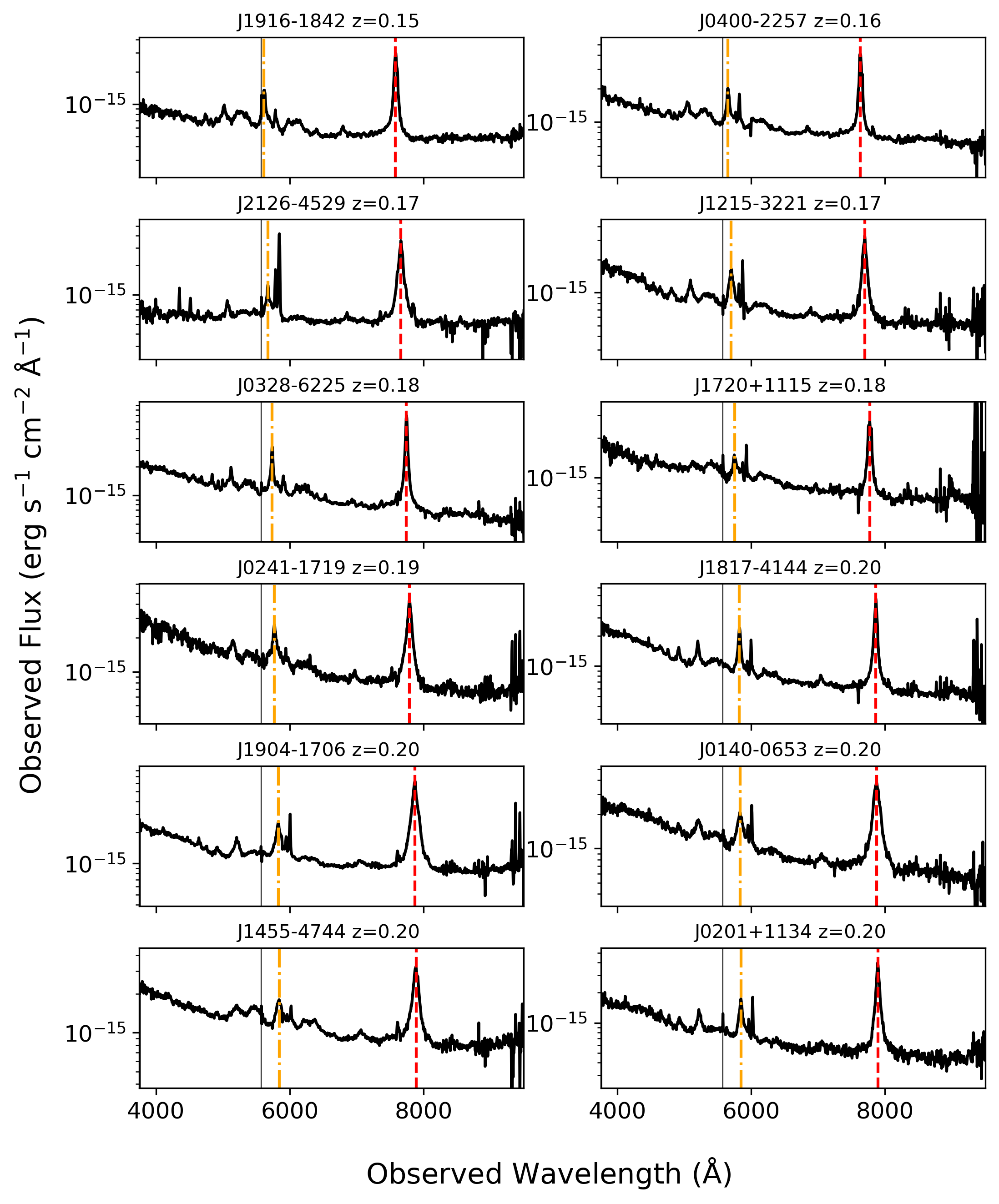}
\caption{As in Fig.~\ref{fig:gallery1}.}
\end{figure*}

\begin{figure*}[t]
\centering
\includegraphics[width=\textwidth]{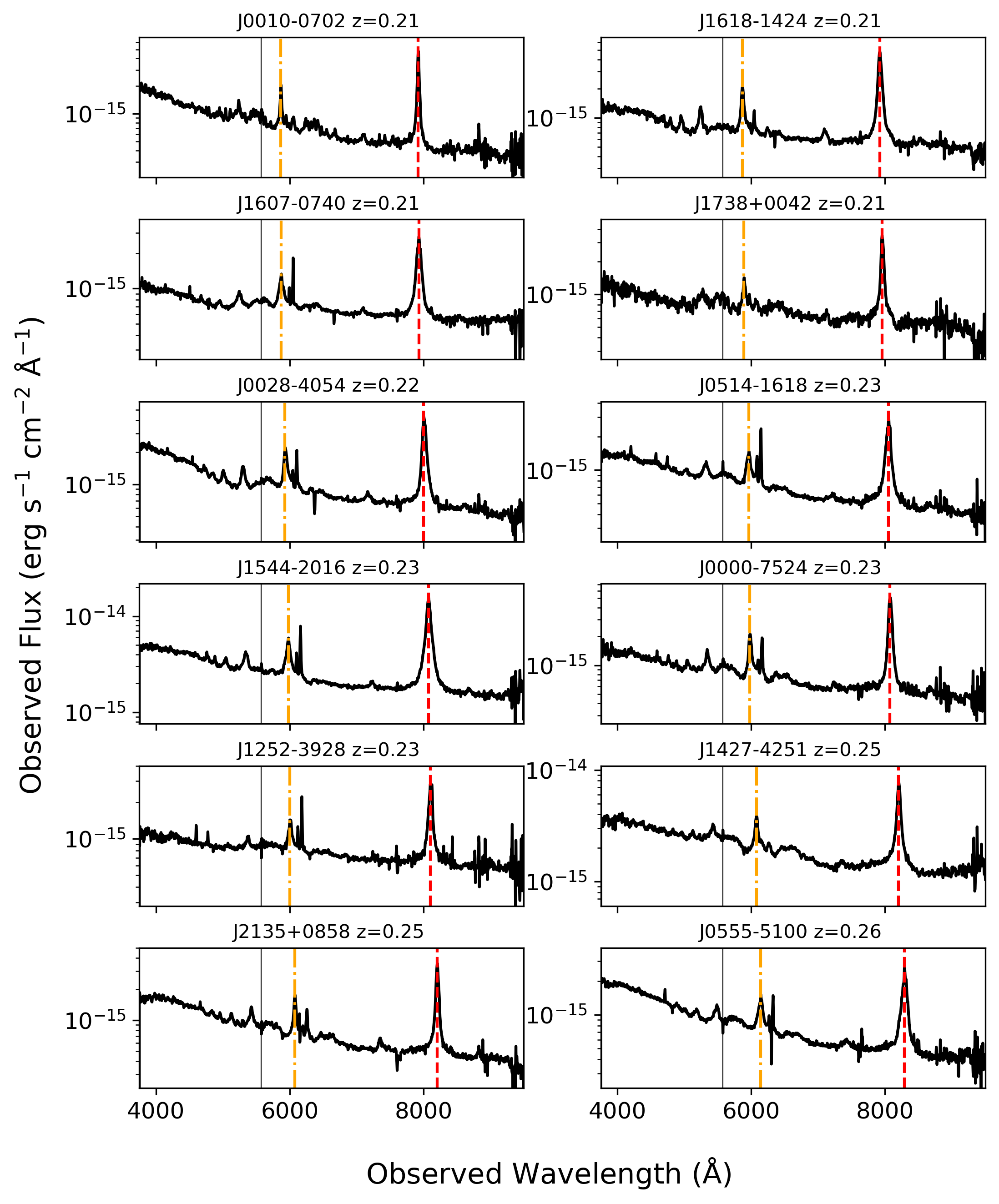}
\caption{As in Fig.~\ref{fig:gallery1}.}
\end{figure*}

\begin{figure*}[t]
\centering
\includegraphics[width=\textwidth]{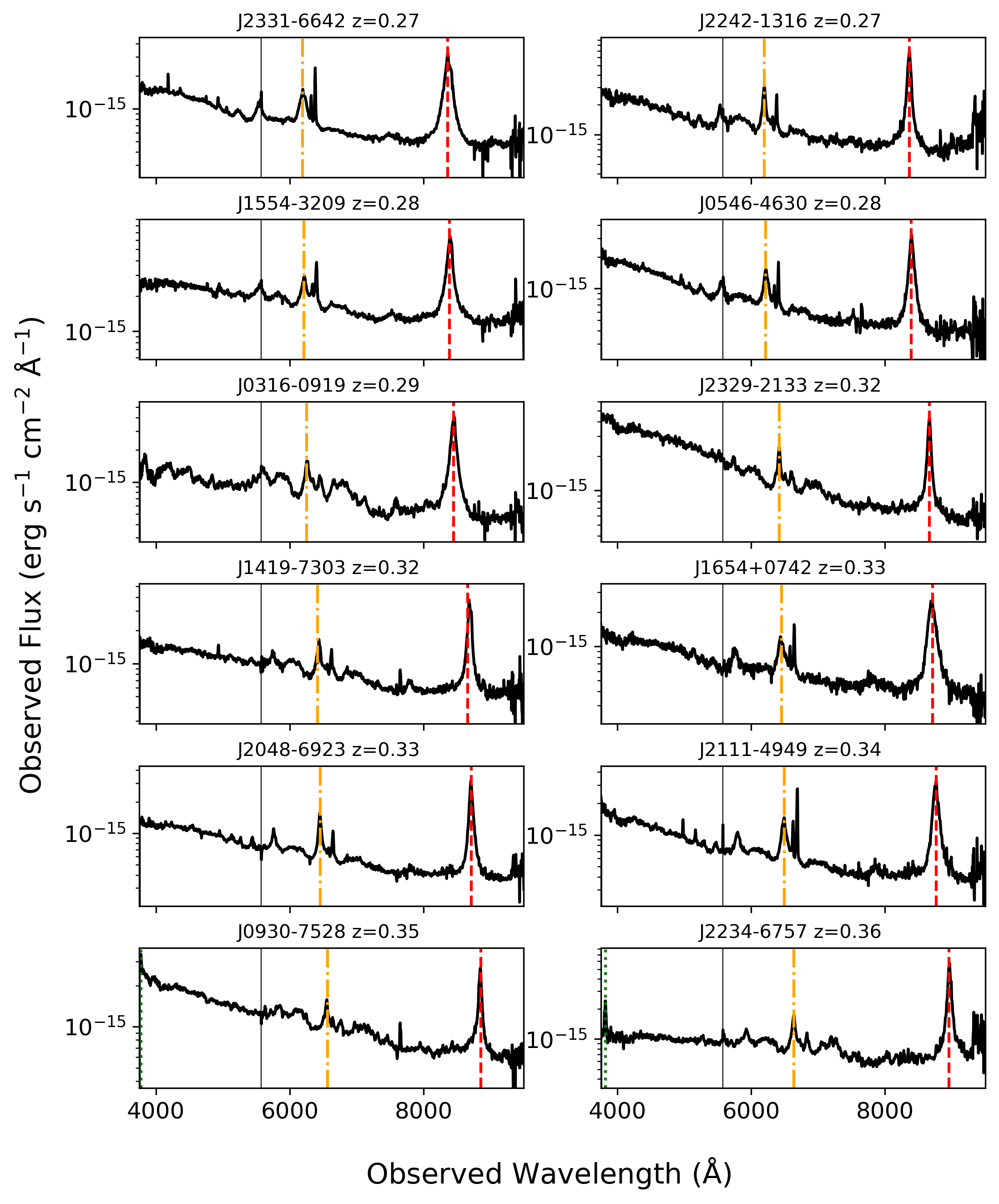}
\caption{As in Fig.~\ref{fig:gallery1}.}
\end{figure*}

\begin{figure*}[t]
\centering
\includegraphics[width=\textwidth]{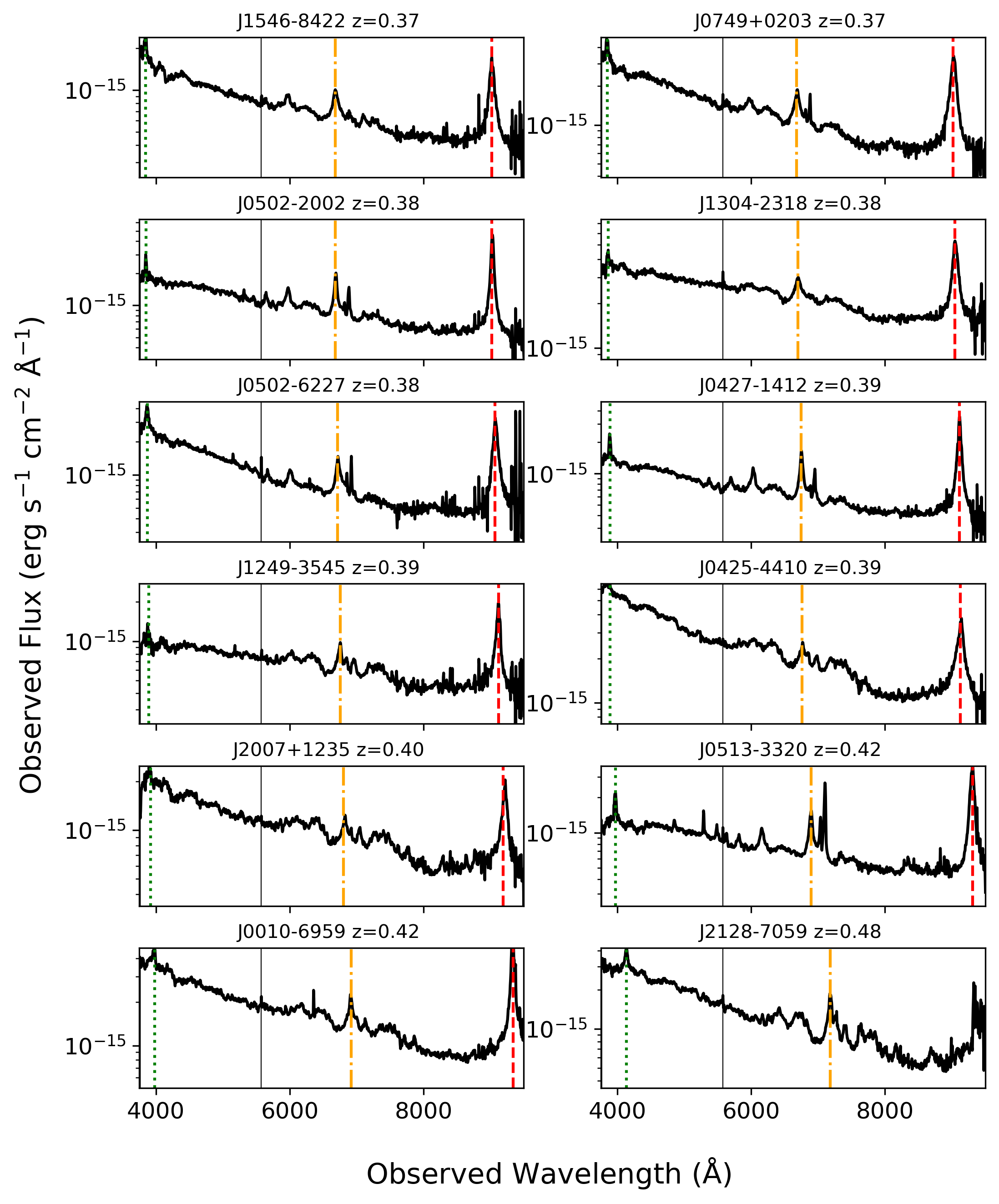}
\caption{As in Fig.~\ref{fig:gallery1}.}
\end{figure*}

\begin{figure*}[t]
\centering
\includegraphics[width=\textwidth]{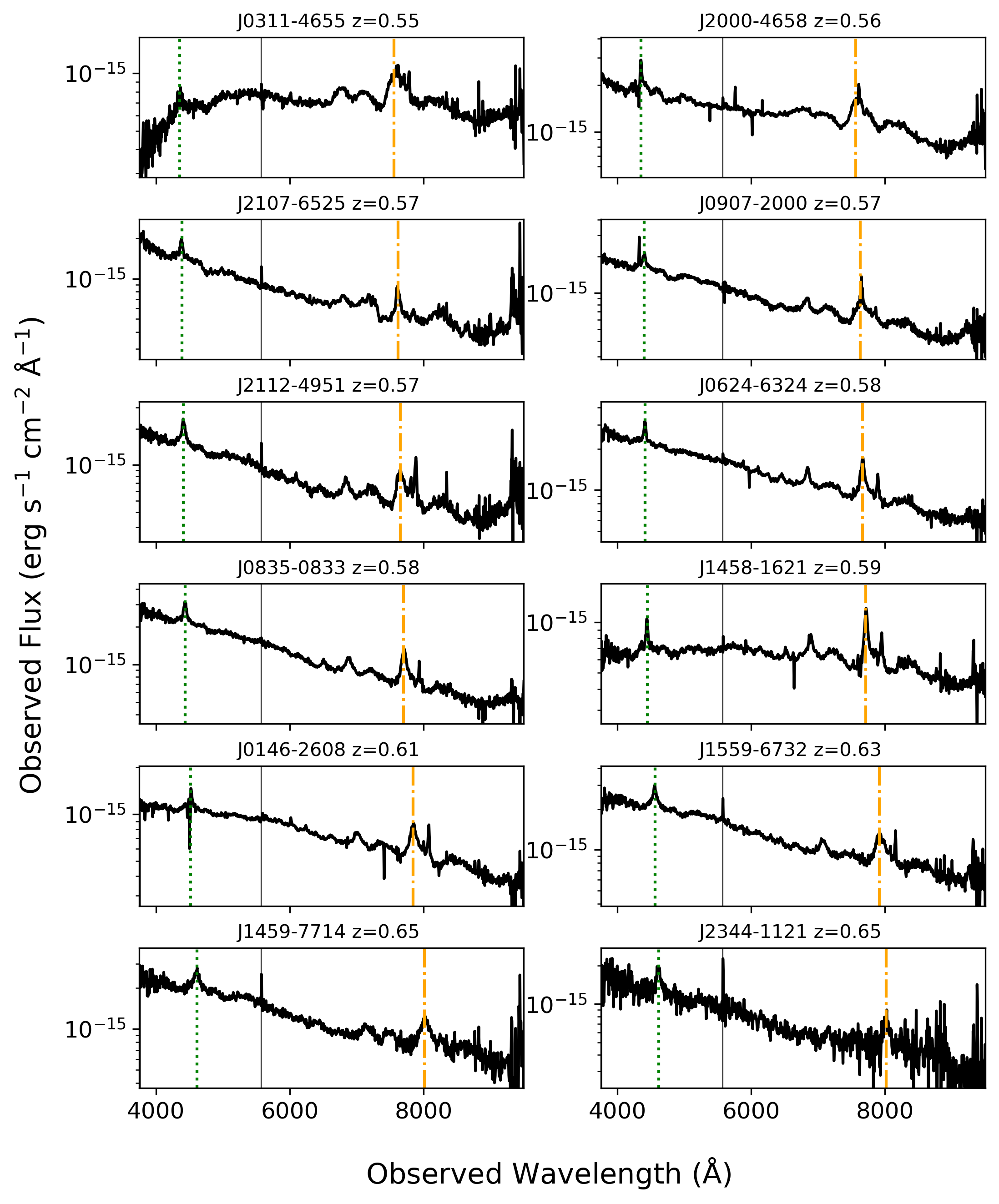}
\caption{As in Fig.~\ref{fig:gallery1}.}
\end{figure*}

\begin{figure*}[t]
\centering
\includegraphics[width=\textwidth]{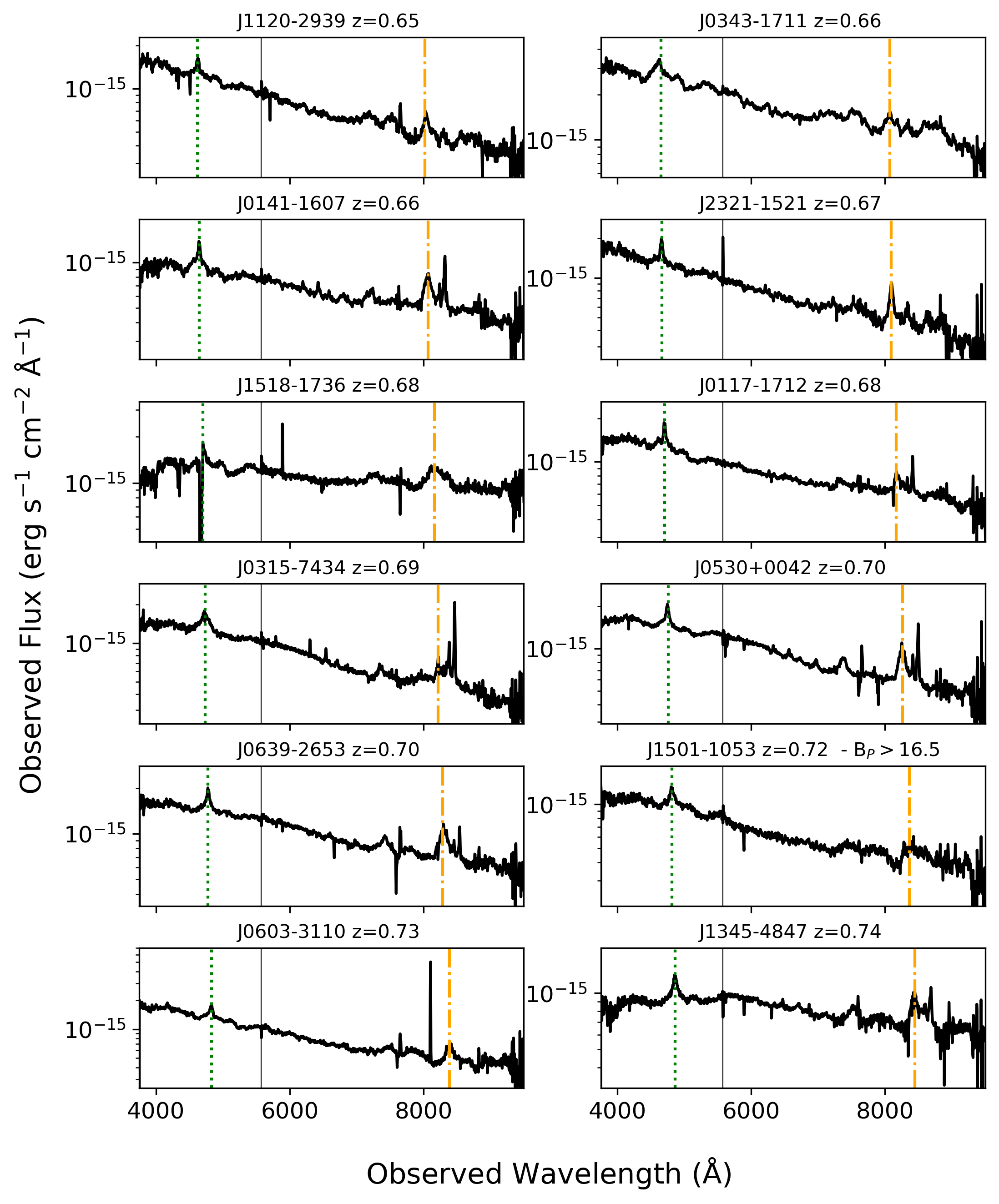}
\caption{As in Fig.~\ref{fig:gallery1}.}
\end{figure*}

\begin{figure*}[t]
\centering
\includegraphics[width=\textwidth]{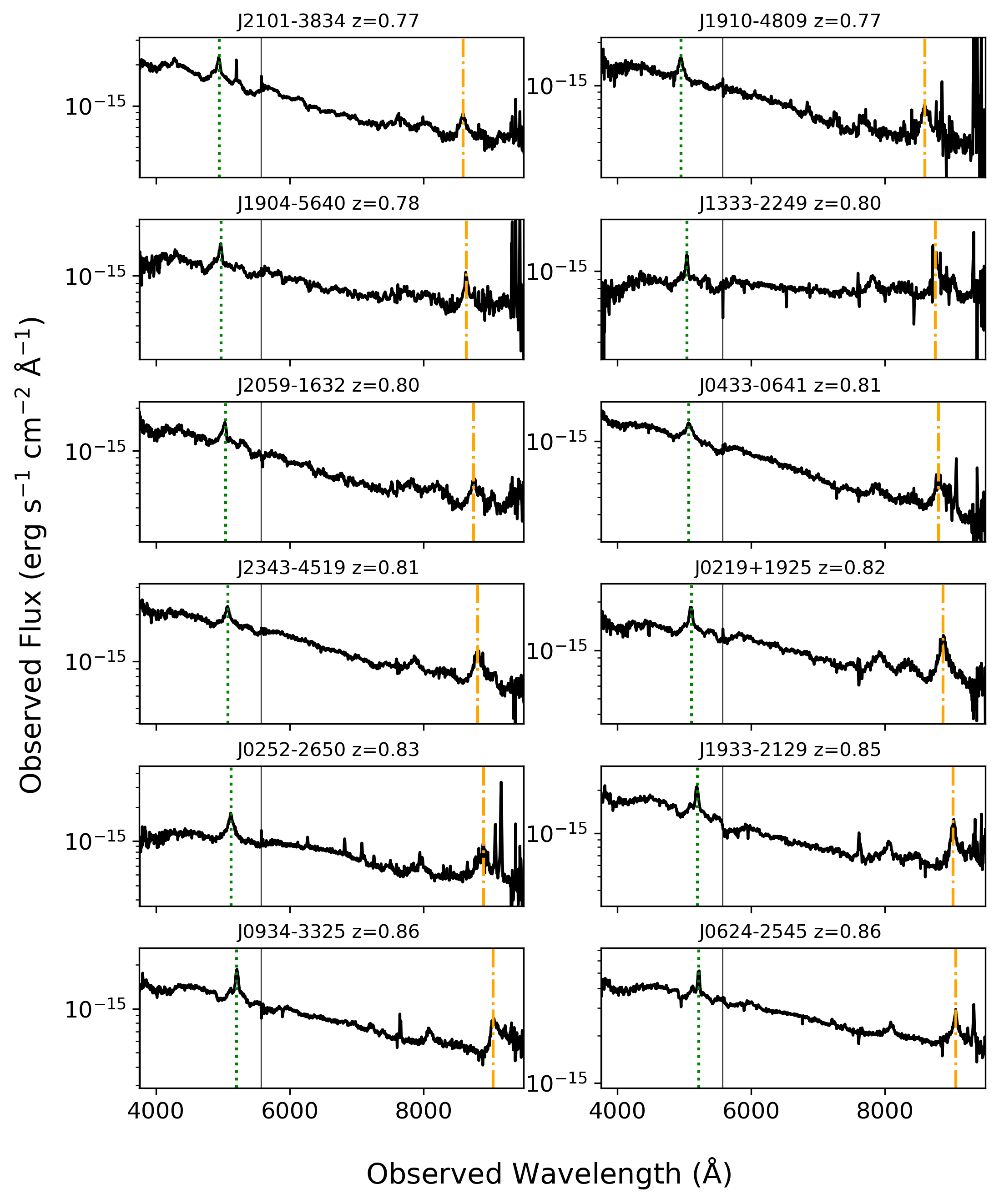}
\caption{As in Fig.~\ref{fig:gallery1}.}
\end{figure*}

\begin{figure*}[t]
\centering
\includegraphics[width=\textwidth]{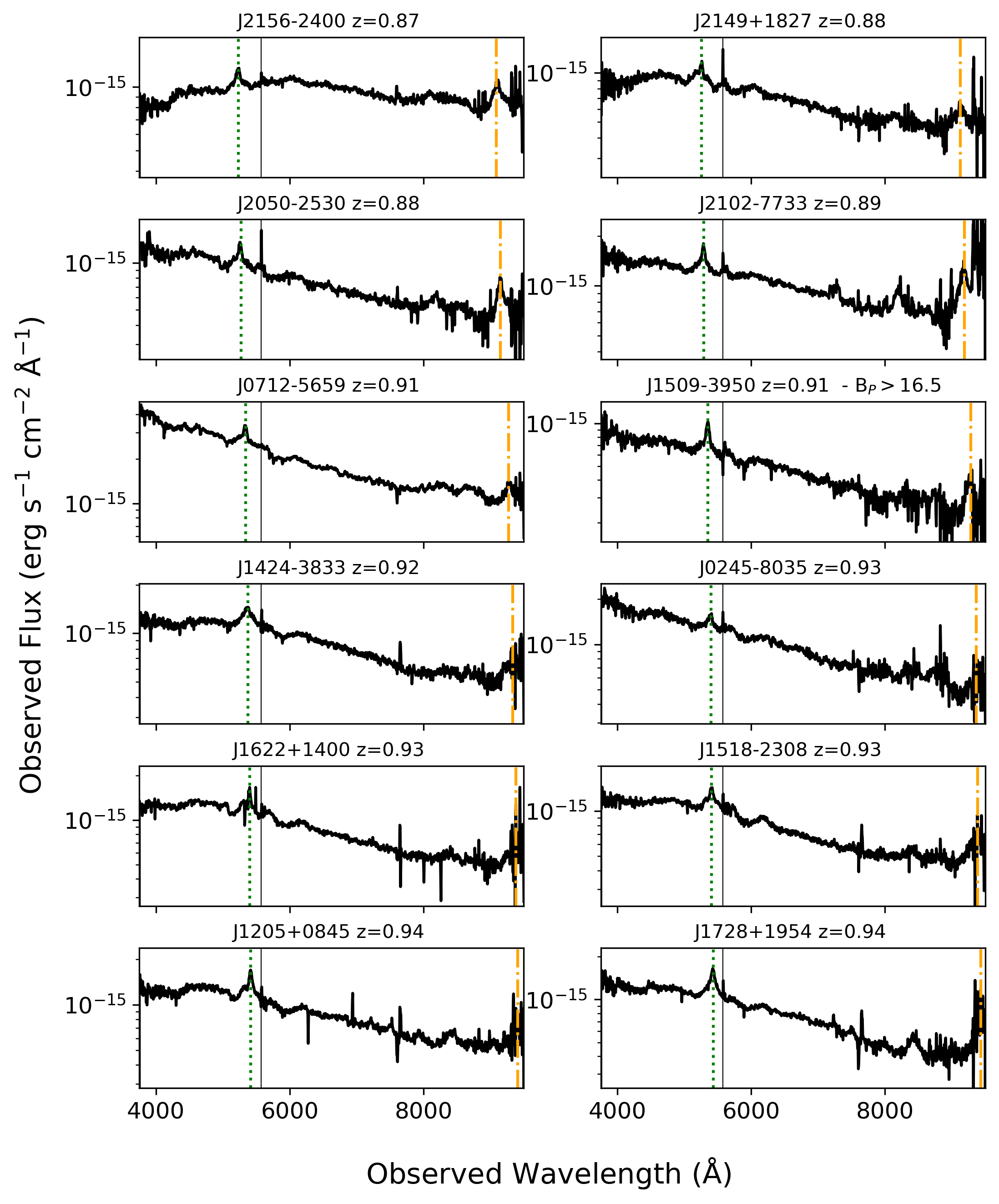}
\caption{As in Fig.~\ref{fig:gallery1}.}
\end{figure*}

\begin{figure*}[t]
\centering
\includegraphics[width=\textwidth]{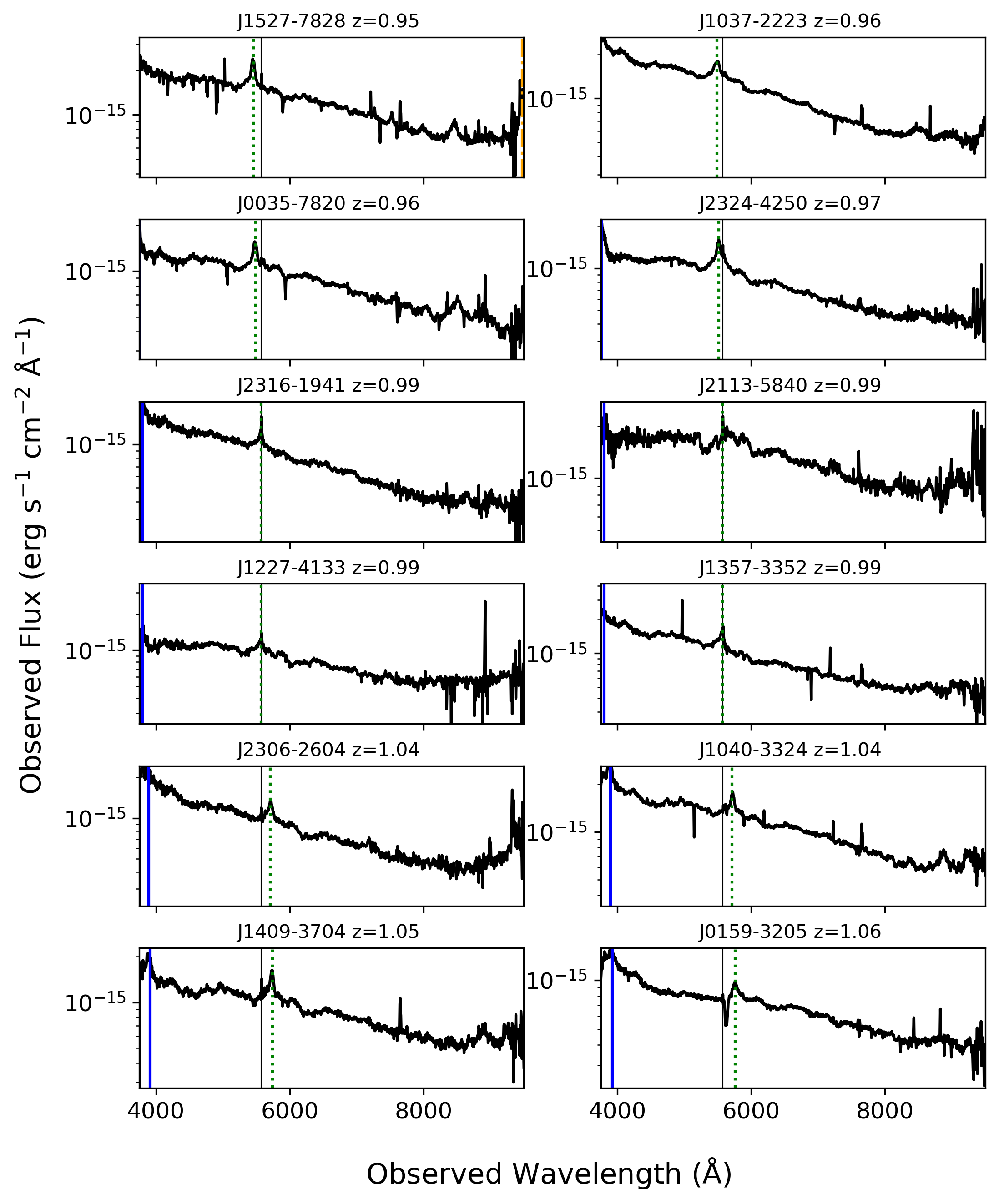}
\caption{As in Fig.~\ref{fig:gallery1}.}
\end{figure*}

\begin{figure*}[t]
\centering
\includegraphics[width=\textwidth]{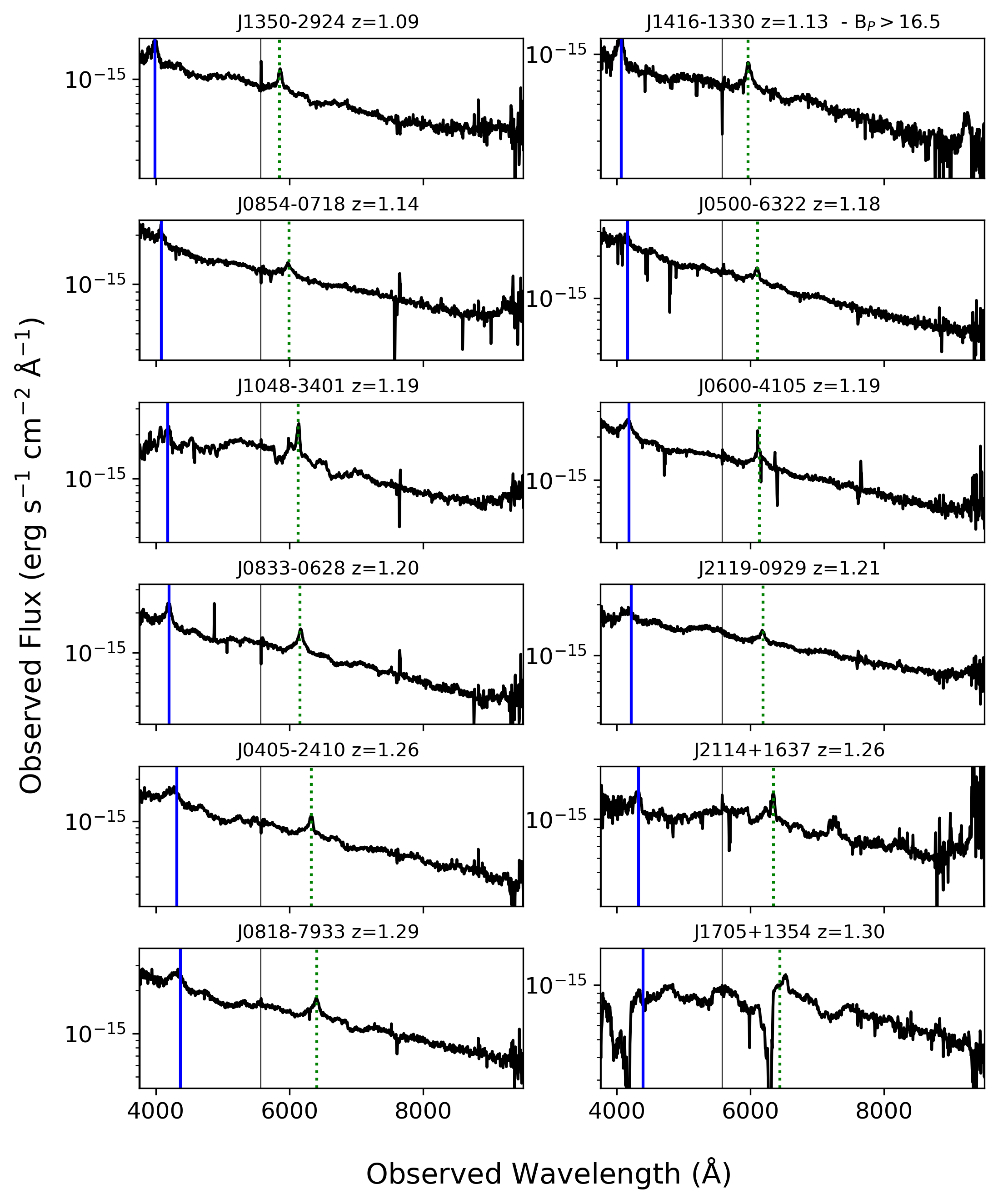}
\caption{As in Fig.~\ref{fig:gallery1}.}
\end{figure*}

\begin{figure*}[t]
\centering
\includegraphics[width=\textwidth]{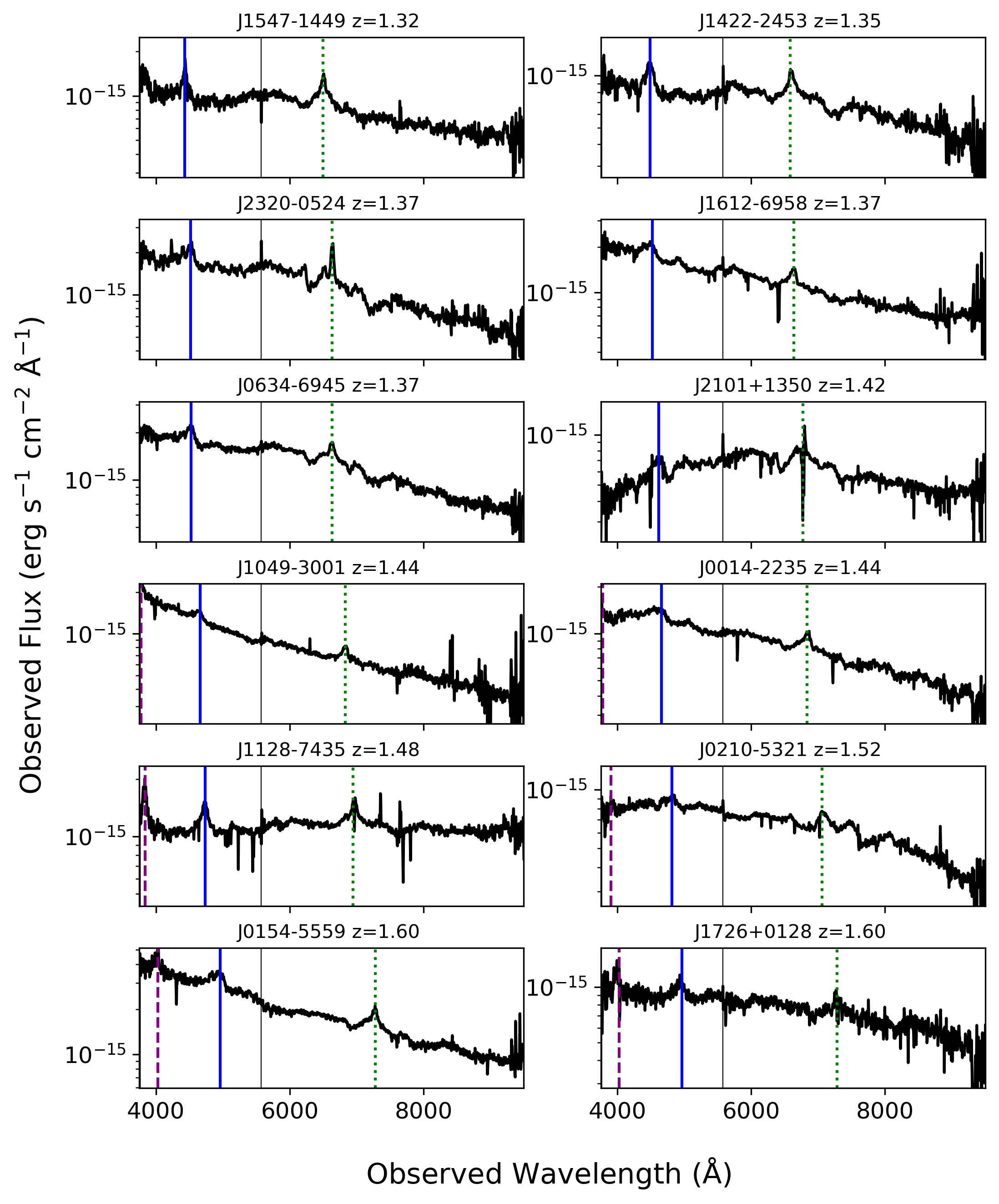}
\caption{As in Fig.~\ref{fig:gallery1}.}
\end{figure*}

\begin{figure*}[t]
\centering
\includegraphics[width=\textwidth]{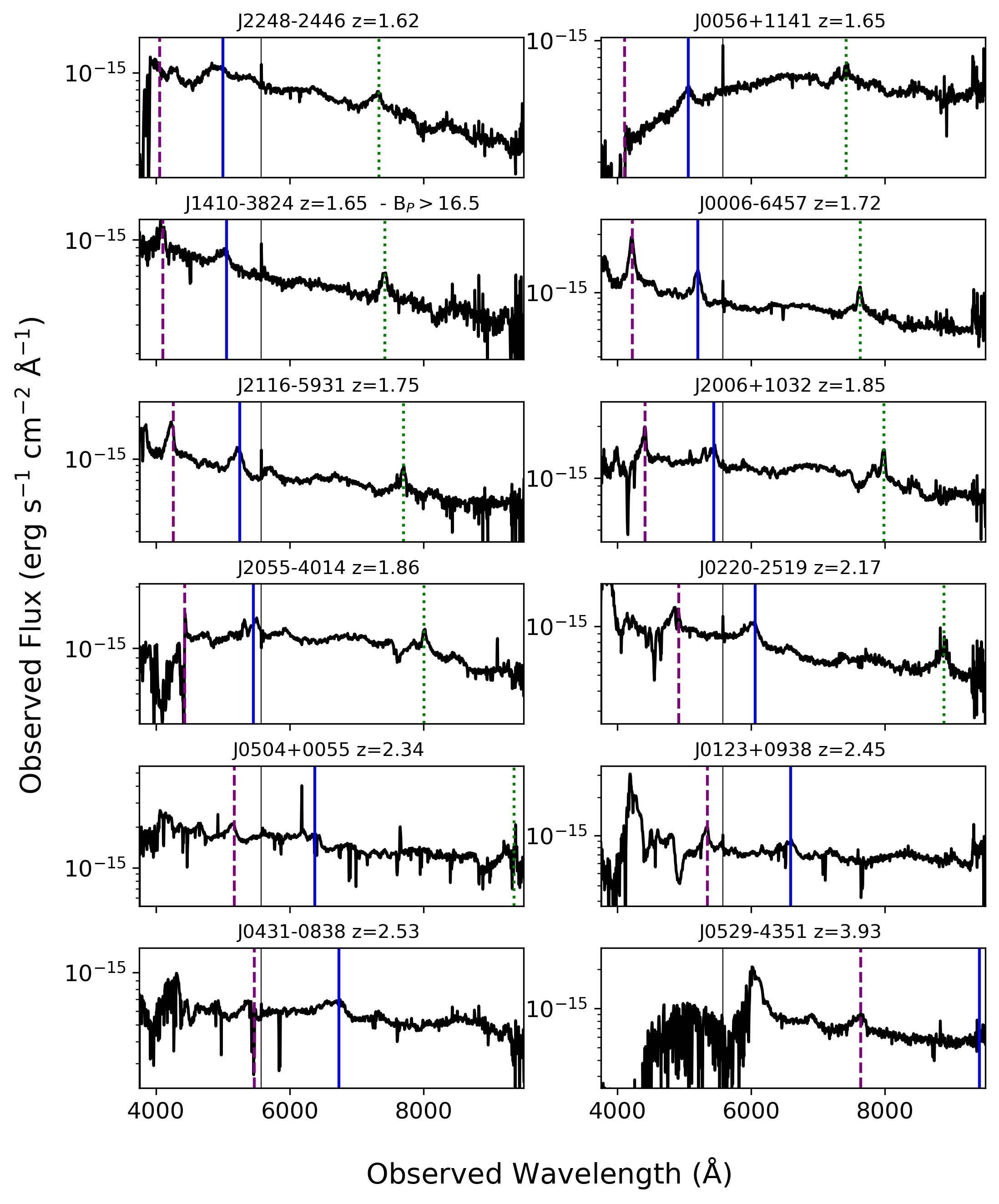}
\caption{As in Fig.~\ref{fig:gallery1}.}
\label{fig:galleryLast}
\end{figure*}

\section{Unclassified sources observed in AllBRICQS}\label{app:unknown}

\begin{figure*}[t]
\centering
\includegraphics[width=\textwidth]{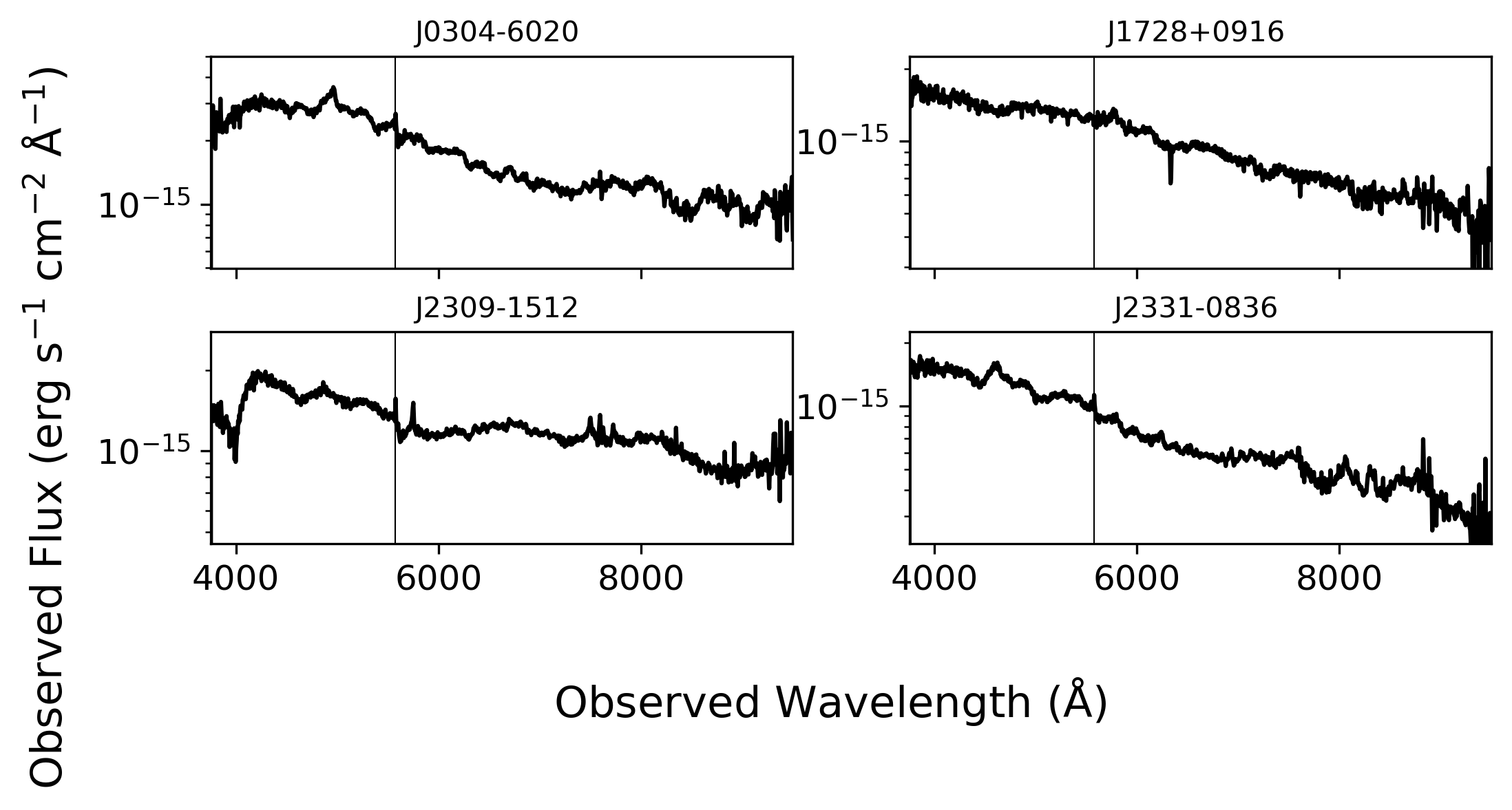}
\caption{WiFeS spectra of the unclassified sources observed in AllBRICQS. The y-axis shows the flux on a logarithmic scale, normalised using their Gaia DR3 photometry. The thin, black vertical line at 5575~\AA\ indicates where the blue and red arms of the spectrograph are spliced together.}
\label{fig:unknown}
\end{figure*}

Table~\ref{tab:unknown} presents the sources which remain unclassified. Column definitions as in \ref{app:sample}. We also present the WiFeS spectra for the unclassified sources in Figure~\ref{fig:unknown}.

\begin{table*}[!ht]
\begin{threeparttable}
\caption{Unidentified AllBRICQS sources}\label{tab:unknown}
\begin{tabular}{llrrcccccl}
\toprule
Name & {\it Gaia} DR3 {\sc source\_id} & RA (J2000) & Dec (J2000) & $B_{P}$ & $R_{P}$ & $W1-W2$ & UVQS & {\it Gaia} & Notes \\
 &  & deg & deg & mag & mag & mag & & & \\
\midrule
J0304-6020 & 4723537966228181376 & 46.1515 & -60.3482 & 15.62 & 15.11 & 1.21 & {--} & GD & \\ % 
J1728+0916 & 4490941985090542720 & 262.2476 & 9.2697 & 16.30 & 15.68 & 1.25 & {--} & GD & \\ % 
J2309-1512 & 2410041905112464256 & 347.4918 & -15.2019 & 16.21 & 15.18 & 1.28 & {--} & G{--} & Possible $z=0.54$\\ % 
J2331-0836 & 2437563029048114176 & 352.8658 & -8.6121 & 16.47 & 16.12 & 1.12 & {--} & GD & Possible $z=0.66$ \\ % 
\bottomrule
\multicolumn{2}{l}{Position information from Gaia DR3.}\\
\multicolumn{10}{l}{Column definitions as in Table~\ref{tab:sample}.}\\
\end{tabular}
\end{threeparttable}
\end{table*}

\section{Stars observed in AllBRICQS}\label{app:nonqso}

\begin{figure*}[t]
\centering
\includegraphics[width=\textwidth]{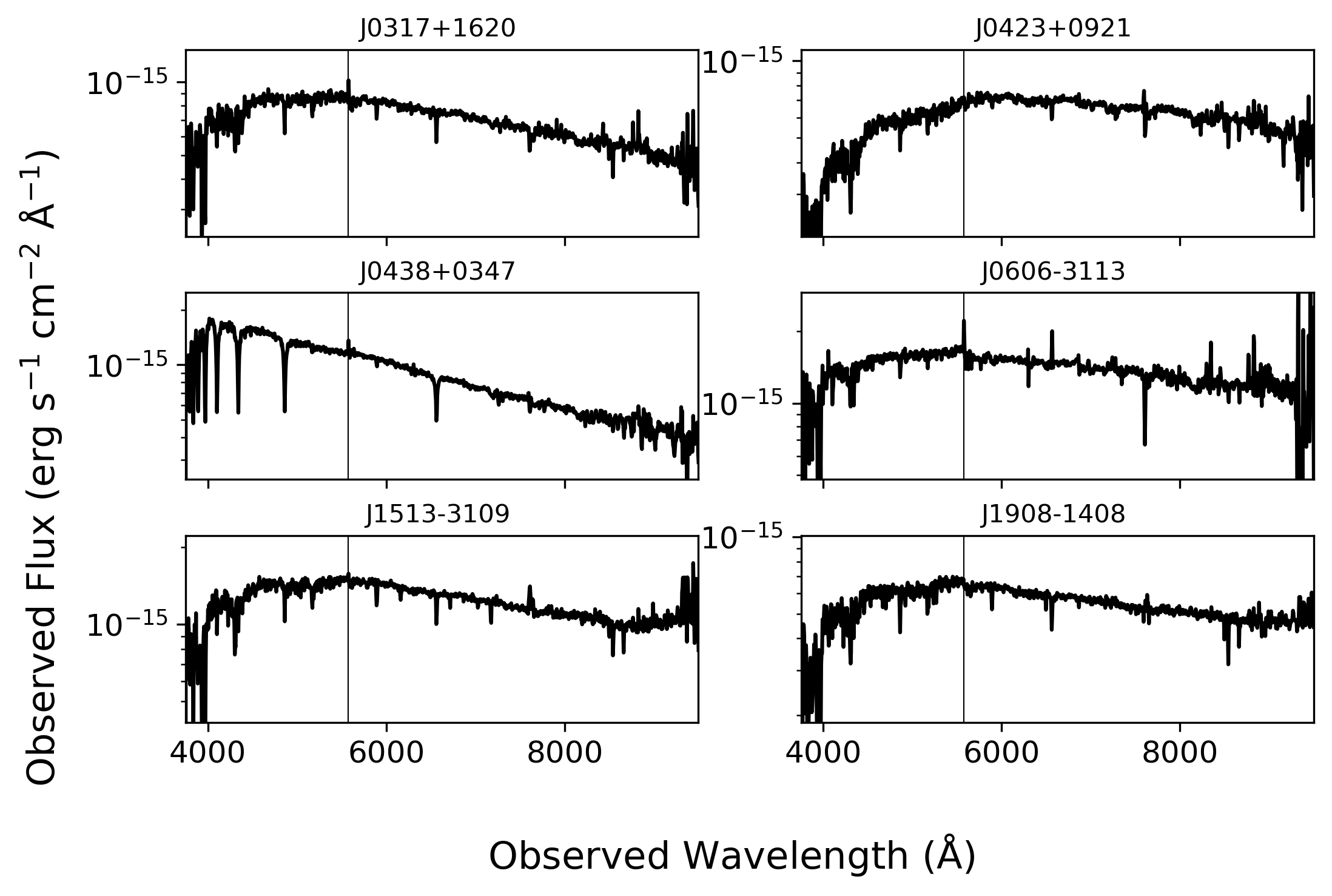}
\caption{WiFeS spectra of the stars observed in AllBRICQS. The y-axis shows the flux on a logarithmic scale, normalised using their Gaia DR3 photometry. The thin, black vertical line at 5575~\AA\ indicates where the blue and red arms of the spectrograph are spliced together.}
\label{fig:stars}
\end{figure*}

In Table~\ref{tab:nonqso}, we present a list of the stars observed in AllBRICQS. Column definitions as in \ref{app:sample}. We also present the WiFeS spectra for the stars in Figure~\ref{fig:stars}.

\begin{table*}[!ht]
\begin{threeparttable}
\caption{AllBRICQS stars}\label{tab:nonqso}
\begin{tabular}{llrrcccccl}
\toprule
Name & {\it Gaia} DR3 {\sc source\_id} & RA (J2000) & Dec (J2000) & $B_{P}$ & $R_{P}$ & $W1-W2$ & UVQS & {\it Gaia} & Notes \\
 &  & deg & deg & mag & mag & mag & & & \\
\midrule
J0317+1620 & 55225724579988736 & 49.4853 & 16.3446 & 16.77 & 15.81 & 0.38 & {--} & {--}{--} & \\ % 
J0423+0921 & 3299580282664334848 & 65.9506 & 9.3590 & 17.07 & 15.80 & 0.33 & {--} & {--}{--} & \\ % 
J0438+0347 & 3281266404675233280 & 69.6230 & 3.7839 & 16.34 & 15.83 & 0.25 & {--} & v{--} & \\ % 
J0606-3113 & 2896198393037594112 & 91.6284 & -31.2207 & 16.07 & 15.04 & 1.72 & {--} & G{--} & \\ % 
J1513-3109 & 6210717877628312448 & 228.4751 & -31.1584 & 16.22 & 15.13 & 0.30 & {--} & {--}{--} & \\ % 
J1908-1408 & 4197369116320824064 & 287.0005 & -14.1379 & 17.12 & 15.99 & 0.51 & {--} & {--}{--} & \\ % 
\bottomrule
\multicolumn{2}{l}{Position information from Gaia DR3.}\\
\multicolumn{10}{l}{Column definitions as in Table~\ref{tab:sample}.}\\
\end{tabular}
\end{threeparttable}
\end{table*}

\section{AllBRICQS Luminosities}\label{app:lum}

In Table~\ref{tab:lum}, we list the continuum and bolometric luminosities for the confirmed AllBRICQS quasars, arranged in order of ascending RA.

\begin{table*}[t]
\begin{threeparttable}
\caption{AllBRICQS Quasar Luminosities}\label{tab:lum}
\begin{tabular}{lcccc|lcccc}
\toprule 
Name & $\log L(1450\text{\AA})$ & $\log L(3000\text{\AA})$ & $\log L(5100\text{\AA})$ & $\log L_{\text{bol}}$ & Name & $\log L(1450\text{\AA})$ & $\log L(3000\text{\AA})$ & $\log L(5100\text{\AA})$ & $\log L_{\text{bol}}$ \\
 & erg~s$^{-1}$ & erg~s$^{-1}$ & erg~s$^{-1}$ & erg~s$^{-1}$ &  & erg~s$^{-1}$ & erg~s$^{-1}$ & erg~s$^{-1}$ & erg~s$^{-1}$ \\
\midrule 
J0000-7524 & \ldots & \ldots & 44.86 & 45.68 & J0327-7224 & \ldots & \ldots & 44.51 & 45.36 \\
J0006-6457 & 47.02 & 46.99 & \ldots & 47.48 & J0328-6225 & \ldots & \ldots & 44.81 & 45.63 \\
J0010-0702 & \ldots & \ldots & 44.72 & 45.55 & J0343-1711 & \ldots & 46.54 & 46.34 & 47.07 \\
J0010-6959 & \ldots & 45.96 & 45.78 & 46.53 & J0400-2257 & \ldots & \ldots & 44.64 & 45.48 \\
J0014-2235 & \ldots & 46.83 & \ldots & 47.39 & J0405-2410 & \ldots & 46.68 & \ldots & 47.24 \\
J0028-4054 & \ldots & \ldots & 44.88 & 45.70 & J0425-4410 & \ldots & 46.06 & 45.81 & 46.59 \\
J0035-7820 & \ldots & 46.50 & \ldots & 47.06 & J0427-1412 & \ldots & 45.51 & 45.33 & 46.10 \\
J0056+1141 & \ldots & 46.93 & \ldots & 47.48 & J0431-0838 & 47.31 & \ldots & \ldots & 47.67 \\
J0117-1712 & \ldots & 46.09 & 46.01 & 46.70 & J0433-0641 & \ldots & 46.24 & 45.93 & 46.74 \\
J0123+0938 & 47.27 & \ldots & \ldots & 47.64 & J0500-6322 & \ldots & 46.79 & \ldots & 47.35 \\
J0140-0653 & \ldots & \ldots & 44.85 & 45.67 & J0502-2002 & \ldots & 45.56 & 45.42 & 46.17 \\
J0141-1607 & \ldots & 45.90 & 45.89 & 46.55 & J0502-6227 & \ldots & 45.69 & 45.34 & 46.20 \\
J0146-2608 & \ldots & 45.90 & 45.79 & 46.50 & J0504+0055 & 47.66 & \ldots & \ldots & 47.99 \\
J0154-5559 & 47.45 & 47.22 & \ldots & 47.78 & J0513-3320 & \ldots & 45.49 & 45.42 & 46.14 \\
J0159-3205 & \ldots & 46.45 & \ldots & 47.02 & J0514-1618 & \ldots & \ldots & 44.88 & 45.70 \\
J0201+1134 & \ldots & \ldots & 44.81 & 45.63 & J0529-4351 & 47.88 & \ldots & \ldots & 48.19 \\
J0207-2354 & \ldots & \ldots & 44.50 & 45.35 & J0530+0042 & \ldots & 46.32 & 46.06 & 46.83 \\
J0210-5321 & \ldots & 46.86 & \ldots & 47.42 & J0533-1434 & \ldots & \ldots & 44.57 & 45.41 \\
J0219+1925 & \ldots & 46.58 & 46.33 & 47.09 & J0546-4630 & \ldots & 45.35 & 45.03 & 45.89 \\
J0220-2519 & \ldots & 47.15 & \ldots & 47.70 & J0555-5100 & \ldots & 45.31 & 44.99 & 45.85 \\
J0241-1719 & \ldots & \ldots & 44.80 & 45.62 & J0600-4105 & \ldots & 46.82 & \ldots & 47.38 \\
J0245-8035 & \ldots & 46.58 & \ldots & 47.14 & J0603-3110 & \ldots & 46.20 & 45.94 & 46.72 \\
J0252-2650 & \ldots & 46.25 & 46.14 & 46.83 & J0624-2545 & \ldots & 46.90 & 46.79 & 47.45 \\
J0311-4655 & \ldots & 45.62 & 45.83 & 46.42 & J0624-6324 & \ldots & 46.19 & 45.96 & 46.72 \\
J0315-7434 & \ldots & 46.16 & 45.97 & 46.71 & J0634-6945 & \ldots & 47.00 & \ldots & 47.55 \\
J0316-0919 & \ldots & 45.28 & 45.14 & 45.91 & J0639-2653 & \ldots & 46.33 & 46.15 & 46.88 \\
\bottomrule
\end{tabular}
\end{threeparttable}
\end{table*}

\setcounter{table}{3}
\begin{table*}[t]
\begin{threeparttable}
\caption{AllBRICQS Quasar Luminosities (cont.)}
\begin{tabular}{lcccc|lcccc}
\toprule 
Name & $\log L(1450\text{\AA})$ & $\log L(3000\text{\AA})$ & $\log L(5100\text{\AA})$ & $\log L_{\text{bol}}$ & Name & $\log L(1450\text{\AA})$ & $\log L(3000\text{\AA})$ & $\log L(5100\text{\AA})$ & $\log L_{\text{bol}}$ \\
 & erg~s$^{-1}$ & erg~s$^{-1}$ & erg~s$^{-1}$ & erg~s$^{-1}$ &  & erg~s$^{-1}$ & erg~s$^{-1}$ & erg~s$^{-1}$ & erg~s$^{-1}$ \\
\midrule 
J0712-5659 & \ldots & 46.84 & \ldots & 47.40 & J1409-3704 & \ldots & 46.60 & \ldots & 47.16 \\
J0715-4951 & \ldots & \ldots & 44.99 & 45.80 & J1410-3824 & 46.86 & 46.79 & \ldots & 47.31 \\
J0749+0203 & \ldots & 45.75 & 45.52 & 46.31 & J1416-1330 & \ldots & 46.49 & \ldots & 47.05 \\
J0818-7933 & \ldots & 46.95 & \ldots & 47.50 & J1419-7303 & \ldots & 45.55 & 45.38 & 46.15 \\
J0833-0628 & \ldots & 46.72 & \ldots & 47.28 & J1422-2453 & \ldots & 46.80 & \ldots & 47.36 \\
J0835-0833 & \ldots & 46.16 & 45.88 & 46.68 & J1424-3833 & \ldots & 46.48 & \ldots & 47.05 \\
J0854-0718 & \ldots & 46.71 & \ldots & 47.27 & J1427-4251 & \ldots & 45.66 & 45.47 & 46.24 \\
J0907-2000 & \ldots & 46.17 & 45.88 & 46.68 & J1455-4744 & \ldots & \ldots & 45.11 & 45.91 \\
J0930-7528 & \ldots & 45.71 & 45.51 & 46.28 & J1458-1621 & \ldots & 45.80 & 45.85 & 46.49 \\
J0934-3325 & \ldots & 46.48 & 46.37 & 47.05 & J1459-7714 & \ldots & 46.39 & 46.11 & 46.89 \\
J1037-2223 & \ldots & 46.59 & \ldots & 47.15 & J1501-1053 & \ldots & 46.17 & 46.02 & 46.74 \\
J1040-3324 & \ldots & 46.70 & \ldots & 47.26 & J1509-3950 & \ldots & 46.23 & \ldots & 46.80 \\
J1048-3401 & \ldots & 46.83 & \ldots & 47.39 & J1518-1736 & \ldots & 46.25 & 46.27 & 46.90 \\
J1049-3001 & \ldots & 46.77 & \ldots & 47.33 & J1518-2308 & \ldots & 46.52 & \ldots & 47.08 \\
J1120-2939 & \ldots & 46.08 & 45.82 & 46.61 & J1527-7828 & \ldots & 46.70 & \ldots & 47.26 \\
J1128-7435 & \ldots & 47.34 & \ldots & 47.88 & J1538-4004 & \ldots & \ldots & 44.61 & 45.45 \\
J1205+0845 & \ldots & 46.43 & \ldots & 47.00 & J1544-2016 & \ldots & \ldots & 45.44 & 46.21 \\
J1215-3221 & \ldots & \ldots & 44.63 & 45.47 & J1546-8422 & \ldots & 45.61 & 45.37 & 46.17 \\
J1227-4133 & \ldots & 46.49 & \ldots & 47.05 & J1547-1449 & \ldots & 46.86 & \ldots & 47.42 \\
J1249-3545 & \ldots & 45.40 & 45.37 & 46.07 & J1554-3209 & \ldots & 45.66 & 45.53 & 46.27 \\
J1252-3928 & \ldots & \ldots & 44.97 & 45.78 & J1559-6732 & \ldots & 46.31 & 46.06 & 46.83 \\
J1304-2318 & \ldots & 46.06 & 45.96 & 46.66 & J1601-7202 & \ldots & \ldots & 44.54 & 45.39 \\
J1333-2249 & \ldots & 46.22 & 46.33 & 46.92 & J1607-0740 & \ldots & \ldots & 44.88 & 45.70 \\
J1345-4847 & \ldots & 46.23 & 46.19 & 46.85 & J1612-6958 & \ldots & 46.97 & \ldots & 47.52 \\
J1350-2924 & \ldots & 46.55 & \ldots & 47.11 & J1618-1424 & \ldots & \ldots & 45.03 & 45.83 \\
J1357-3352 & \ldots & 46.52 & \ldots & 47.08 & J1618-3059 & \ldots & \ldots & 44.16 & 45.04 \\
\bottomrule
\end{tabular}
\end{threeparttable}
\end{table*}

\setcounter{table}{3}
\begin{table*}[t]
\begin{threeparttable}
\caption{AllBRICQS Quasar Luminosities (cont.)}
\begin{tabular}{lcccc|lcccc}
\toprule 
Name & $\log L(1450\text{\AA})$ & $\log L(3000\text{\AA})$ & $\log L(5100\text{\AA})$ & $\log L_{\text{bol}}$ & Name & $\log L(1450\text{\AA})$ & $\log L(3000\text{\AA})$ & $\log L(5100\text{\AA})$ & $\log L_{\text{bol}}$ \\
 & erg~s$^{-1}$ & erg~s$^{-1}$ & erg~s$^{-1}$ & erg~s$^{-1}$ &  & erg~s$^{-1}$ & erg~s$^{-1}$ & erg~s$^{-1}$ & erg~s$^{-1}$ \\
\midrule 
J1619-7832 & \ldots & \ldots & 44.63 & 45.47 & J2102-7733 & \ldots & 46.58 & \ldots & 47.14 \\
J1622+1400 & \ldots & 46.46 & \ldots & 47.03 & J2107-6525 & \ldots & 45.90 & 45.72 & 46.47 \\
J1654+0742 & \ldots & 45.40 & 45.21 & 45.99 & J2111-4949 & \ldots & 45.35 & 45.14 & 45.94 \\
J1705+1354 & \ldots & 46.85 & \ldots & 47.41 & J2112-4951 & \ldots & 45.95 & 45.68 & 46.48 \\
J1720+1115 & \ldots & \ldots & 44.91 & 45.72 & J2113-5840 & \ldots & 46.73 & \ldots & 47.29 \\
J1726+0128 & 47.07 & 47.05 & \ldots & 47.53 & J2114+1637 & \ldots & 46.85 & \ldots & 47.41 \\
J1728+1954 & \ldots & 46.46 & \ldots & 47.03 & J2116-5931 & 47.03 & 46.99 & \ldots & 47.48 \\
J1738+0042 & \ldots & \ldots & 45.03 & 45.83 & J2119-0929 & \ldots & 46.85 & \ldots & 47.41 \\
J1817-4144 & \ldots & \ldots & 44.83 & 45.65 & J2126-4529 & \ldots & \ldots & 44.43 & 45.29 \\
J1904-1706 & \ldots & \ldots & 45.11 & 45.91 & J2128-7059 & \ldots & 46.03 & 45.69 & 46.53 \\
J1904-5640 & \ldots & 46.31 & 46.26 & 46.92 & J2135+0858 & \ldots & 45.16 & 44.92 & 45.75 \\
J1905-2639 & \ldots & \ldots & 44.65 & 45.49 & J2149+1827 & \ldots & 46.38 & \ldots & 46.95 \\
J1910-4809 & \ldots & 46.25 & 46.09 & 46.81 & J2156-2400 & \ldots & 46.40 & \ldots & 46.97 \\
J1916-1842 & \ldots & \ldots & 44.45 & 45.30 & J2207-1654 & \ldots & \ldots & 44.24 & 45.11 \\
J1933-2129 & \ldots & 46.48 & 46.42 & 47.07 & J2234-6757 & \ldots & 45.30 & 45.39 & 46.05 \\
J2000-4658 & \ldots & 46.06 & 46.04 & 46.70 & J2242-1316 & \ldots & 45.50 & 45.24 & 46.06 \\
J2006+1032 & \ldots & 47.30 & \ldots & 47.84 & J2248-2446 & \ldots & 46.89 & \ldots & 47.44 \\
J2007+1235 & \ldots & 45.87 & 45.64 & 46.42 & J2306-2604 & \ldots & 46.48 & \ldots & 47.05 \\
J2034-0405 & \ldots & \ldots & 44.32 & 45.19 & J2316-1941 & \ldots & 46.43 & \ldots & 47.00 \\
J2048-6923 & \ldots & 45.30 & 45.14 & 45.91 & J2320-0524 & \ldots & 46.93 & \ldots & 47.48 \\
J2050-2530 & \ldots & 46.34 & \ldots & 46.91 & J2321-1521 & \ldots & 46.09 & 45.84 & 46.62 \\
J2055-4014 & \ldots & 47.28 & \ldots & 47.83 & J2324-4250 & \ldots & 46.41 & \ldots & 46.98 \\
J2058-1452 & \ldots & \ldots & 44.62 & 45.46 & J2329-2133 & \ldots & 45.75 & 45.39 & 46.26 \\
J2059-1632 & \ldots & 46.29 & 46.08 & 46.83 & J2331-6642 & \ldots & 45.20 & 45.02 & 45.81 \\
J2101+1350 & \ldots & 46.81 & \ldots & 47.37 & J2343-4519 & \ldots & 46.45 & 46.30 & 47.00 \\
J2101-3834 & \ldots & 46.33 & 46.15 & 46.88 & J2344-1121 & \ldots & 46.00 & 45.86 & 46.58 \\
\bottomrule
\end{tabular}
\end{threeparttable}
\end{table*}

\end{document}